\documentclass[12pt,preprint]{aastex}
\newcommand{\NeIV}{[Ne\,{\sc iv}]}
\newcommand{\CIV}{C\,{\sc iv}}
\newcommand{\MgII}{Mg\,{\sc ii}}
\newcommand{\FeII}{Fe\,{\sc ii}}
\newcommand{\Hbeta}{H\,{\sc $\beta$}}

\def\mbh{{$\mathcal M_{\textrm{BH}}$}}
\def\msol{{$\mathcal M_{\odot}$}}
\def\Msigma{{$\mathcal M_{\textrm{BH}}-\sigma_{\star}$}}
\def\kmps{{km~s${}^{-1}$}}
\def\ergps{{erg~s${}^{-1}$}}
\def\percm3{{cm${}^{-3}$}}

\shorttitle{MgII-Based Quasar BH Masses}
\shortauthors{Rafiee and Hall}
\begin{document}
\title{Supermassive Black Hole Mass Estimates Using Sloan Digital Sky
Survey Quasar Spectra at $0.7 < z < 2$}
\author{Alireza Rafiee, Patrick B. Hall}
\affil{Department of Physics \& Astronomy, York University,
Toronto, Ontario M3J 1P3, Canada}
%=======================================================================================================
\begin{abstract}
We present \MgII\ -based black hole mass estimates for 27,602
quasars with rest-frame UV spectra available in the Sloan Digital Sky
Survey Data Release Three.  This estimation is possible due to the
existence of an empirical correlation between the radius of the broad line
region and the continuum luminosity at 3000 Angstroms.  We regenerate this
correlation by applying our measurement method to UV spectra of low-redshift
quasars in the HST/IUE databases which have corresponding reverberation
mapping estimates of the \Hbeta\ broad line region's radius.
Our mass estimation method uses the line dispersion rather than the
full width at half maximum of the low-ionization \MgII\ emission line.
We measure \MgII\ line dispersions for quasars
whose spectra have been reconstructed using the most significant eigenspectra
produced through Principal Component Analysis.
We have tested the reliability of using reconstructed spectra in black hole
mass estimation using a Monte Carlo simulation and by comparing the results
from original and reconstructed Data Release Three spectra.  We show that
using reconstructed spectra not only makes bias-free mass estimation
possible for quasars with low spectroscopic signal-to-noise ratio, but also
reduces the intrinsic scatter of the distribution of the black hole masses to
lower than 0.15 dex.
\end{abstract}
%=======================================================================================================
%=======================================================================================================
\section{Introduction} \label{sec:intro}

Over the past decade, we have learned that all large galaxies have supermassive black holes at their centers (e.g., Kormendy \& Richstone et al.~1995; Richstone et al.~1998). Material falling onto black holes forms a rapidly spinning disk which heats up and radiates the light we see as quasars and other active galactic nuclei (AGN) (e.g., Salpeter~1964; Zel'dovich \& Novikov~1964; Lynden Bell~1969).
The masses of these black holes can be measured through different techniques. In galaxies with a dormant black hole, the BH mass can be measured through the velocity dispersion of stars or gas close enough to the BH which their dynamics are dominated by BH gravity (within the radius of influence; e.g., Gebhardt et al.~2000; Ferrarese \& Merritt~2000; Tremaine et al.~2002). However, this method cannot be used for active nuclei where the nucleus outshines the core of the host galaxy (e.g., Richstone et al.~1998; Gebhardt et al.~2000; Ferrarese \& Merritt~2000; Tremaine et al.~2002). Thus, alternative mass measurement methods are necessary, but are AGN type dependent (Blandford \& McKee~1982; Peterson~1993; Miyoshi et al.~1995).
In nearly edge-on (Type 2) AGNs, mass can be measured from water masers located in the outer accretion disk (Miyoshi et al.~1995). However, water masers cannot be used widely due to the extreme edge-on alignment needed for radiation to be sufficiently amplified. In more face-on (Type 1) AGNs, one can instead use reverberation mapping when variations of the broad emission lines relative to the continuum are present (e.g., Koratkar \& Gaskell~1991; Kaspi et al.~2000; Peterson et al.~2004; Kaspi et al.~2005; Bentz et al.~2006; Bentz~2009).

Reverberation mapping (RM) studies (Blandford \& McKee~1982; Peterson~1993) take advantage of long monitoring campaigns of AGNs to infer properties of the broad line emitting region (BLR) through characterization of time delays ($\tau$) between broad emission line and continuum flux variations (e.g., Koratkar \& Gaskell~1991; Kaspi et al.~2000; Kaspi et al.~2005; Bentz et al.~2006; Bentz~2009).
In principle, RM can generate a velocity-delay map (revealing kinematic structure in the BLR) which can help one to estimate the black hole mass independent of other methods (see Done \& Krolik~1996; Ulrich \& Horne~1996; Kollatschny~2003; Horne et al.~2004; Denney et al.~2009; Bentz et al.~2009). However, at the moment, one still has to calibrate quasar black hole masses to nearby well studied dormant black holes (assuming that they are descendants of the active BHs; see, e.g., Kaspi et al.~2000; Krolik~2001; McLure \& Dunlop~2001; Onken et al.~2004).
It has taken many years of RM campaigns to obtain a large sample of such mass measurements even for a handful of low redshift quasars (e.g., Done \& Krolik~1996; Ulrich \& Horne~1996; Kollatschny~2003; Peterson~2006; Bentz et al.~2009; Denney et al.~2009).
The RM results show an empirical correlation between the distance of the BLR from the BH (which is the measured time-lag $\tau$ times the speed of light) and a monochromatic continuum luminosity (known as $R-L$ or $\tau-L$ correlation; Kaspi et al.~2000; Bentz et al.~2006). This $R-L$ correlation simplifies the mass measurement since determining the radius of the BLR using long monitoring campaigns can be replaced by measuring the monochromatic continuum luminosity from a single epoch spectrum.

The RM results also show a relationship between emission line widths ($\Delta V$) and corresponding time-lags (e.g., Peterson \& Wandel~1999, 2000; Onken \& Peterson~2002; Kollatschny~2002) which confirms the virialization of the BLR.
This relationship enables a secondary method of BH mass determination
from $\frac{GM}{R}\propto (\Delta V)^{2}$ if one assumes that the $\Delta V$ of a virialized emission line (e.g., \Hbeta) reflects the virialized bulk motion of the emitting gas (e.g., Peterson \& Wandel~1999, 2000; Onken \& Peterson~2002; Kollatschny~2003).
%---------------
By measuring the line width of a particular line (using the FWHM or the line dispersion) and the luminosity of the AGN at a particular wavelength, one can calibrate the virial mass equation of the secondary method for a sample of RM black hole masses. That equation can then be used to estimate unknown quasar BH masses (e.g., Wandel et al.~1999; Vestergaard~2002; McLure \& Jarvis~2002; Woo \& Urry~2002; Vestergaard \& Peterson~2006; Shen et al.~2008; Wang et al.~2009).

Much work on distant black hole mass estimates has used the hydrogen \Hbeta\ emission line and the 1549~\AA\ doublet of triply ionized carbon (\CIV); e.g., Kaspi et al.~(2000), Vestergaard~(2002), McLure \& Jarvis~(2002) and Vestergaard \& Peterson~(2006). \Hbeta\ can be identified in low redshift quasars ($z\lesssim 0.8$) with observations from ground-based telescopes in the optical or near infrared, while the \CIV\ emission line can be identified in high redshift quasars ($z\gtrsim 2$) in those bands. Thus, \Hbeta\ provides well calibrated masses but cannot be seen in the most distant black holes with ground based telescopes, while \CIV\ can be seen but could be strongly contaminated by non-gravitational motions. For example, the peak of the \CIV\ emission line is usually blueshifted, probably implying an outflow of material (e.g., Leighly et al.~2000; Yuan et al.~2004; Vestergaard~2009).

We have assembled a catalogue of black hole mass estimates using the 2798~\AA\ doublet of singly ionized magnesium (\MgII), which can be observed by ground-based telescopes in the optical or near infrared for even the most distant quasars and which is thought to probe largely the same region of the accretion disk as \Hbeta\ (McLure \& Jarvis~2002; but see Wang et al.~2009). Although there is no RM study which demonstrates the virialization of the \MgII\ emission line, BH mass estimates using \MgII\ show consistency with similar estimates using \Hbeta\ (McLure \& Jarvis~2002), which is considered a virialized line (Peterson \& Wandel~1999).

For our mass estimates, we use the line dispersion of the line profile instead of FWHM since the relation between rms line dispersion and RM time-delay appears to be less scattered than the same relation with rms FWHM, at least for the \Hbeta\ line (Peterson et al. 2004). The line dispersion uses the data directly but is sensitive to the wings of the line profile (Collin et al.~2006).
The FWHM is the FWHM of a fit of one or more Gaussians to the line profile (e.g, McLure \& Dunlop~2004; Shen et al.~2008), and the FWHM of a single Gaussian fit in particular may be a poor representation of the true FWHM of a line.
Furthermore, it is not confirmed that the \MgII\ FWHM is linearly related to the rms FWHM of \Hbeta\ (Wang et al.~2009).
Shen et al.~(2008) have provided a BH mass catalogue with mass estimates made using the FWHMs of \MgII, \Hbeta\ and \CIV. However, they have not calibrated the \MgII\ scaling relation but instead have adopted the McLure \& Dunlop~(2004) relationship, despite there being many updated RM mass measurements since 2004.
Onken \& Kollmeier~(2008) have shown that the Shen et al.~(2008) \MgII\ mass estimates are Eddington ratio dependent, which may indicate a problem with FWHM-based \MgII\ mass estimates.
We are therefore motivated to test the alternative, the line dispersion of \MgII.

To reduce the uncertainties in the line dispersion due to low signal-to-noise ratio spectra and the sensitivity of the technique to the line wings, we have reconstructed the quasar spectra using the Karhunen-Lo\`{e}ve transformation technique (also called Principal Components Analysis; PCA) e.g., Boroson and Green~(1992); Connolly et al.~(1995); Yip et al.~(2004); Boroson et al.~(2010). Reconstructing the quasar spectra reduces the noise sufficiently for us to use the line dispersion robustly.

The outline of this paper is as follows. In \S 2, we describe how we measure the second moment of the \MgII\ line profile, including how we model \FeII\ emission lines and establish a local continuum. In \S 3, we calibrate a mass scaling relation using the \MgII\ line dispersion and the monochromatic continuum luminosity at 3000{\AA}. In \S4, we describe how we select our data sample from the SDSS. In \S 5, we explain how we reduce the noise level by reconstructing the SDSS quasar spectra using PCA. In \S 6, we study the effect of the radiation pressure on BH mass estimates and calibrate a new mass scaling relation for that scenario. In \S 7, we simulate the effects of noise in reconstructed spectra to understand the biases on BH mass estimates before and after reconstruction. In \S 8, we use the mass scaling relations to estimate the BH masses of SDSS quasars and present our BH mass catalogue. We then compare our BH mass estimates with Shen et al.~(2008). In \S 9, we present our conclusions.

In this study we assume a Lambda Cold Dark Matter ($\Lambda CDM$) cosmology with Hubble constant
$H_{0}=71$ km s$^{-1}$ Mpc$^{-1}$, fractional dark energy density $\Omega_{\Lambda}=0.74$ and fractional matter energy density $\Omega_{m}=0.26$ (Spergel et al.~2007) to calculate necessary cosmological quantities.

%=======================================================================================================
\section{Black Hole Mass Estimates via Secondary Method}\label{sec:BHME-WP}

To compute \mbh\ from the virial equation
\begin{equation}\label{equ:virial_equation}
  \mathcal M_{\textrm{BH}} = \frac{\bar{f} R_{BLR}(\Delta V)^2}{G}.
\end{equation}
it is necessary to estimate the radius of the broad line region, $R_{BLR}$, via
a correlation with the AGN continuum luminosity when there is no
reverberation mapping data available. The observed line width, $\Delta V$, and the continuum luminosity can both be estimated from a single epoch spectrum.
The average geometrical factor $\bar{f}$ will remain as a free parameter to be adjusted such that it makes
these secondary mass estimates most consistent with the virial mass estimates.
The virial masses are in turn calibrated to direct mass measurements of nearby inactive black holes through the BH mass - bulge velocity dispersion (\Msigma) relationship (e.g., Gebhardt et al.~2000; Ferrarese \& Merritt~2000; Tremaine et al.~2002). This calibration is valuable since there is no direct way to estimate the mass zero point for the quasar BH mass relationship, but it does assume that the same relationship is valid for quasars and for inactive BHs.

%=======================================================================================================
\subsection{Measuring the Line Width of an Emission Line}\label{sec:measuring line width}
The traditional way to measure the line width is to measure its
\textrm{FWHM}. This is simple for single peaked lines and more complex for
multiple-peaked lines or noisy data (for double-peaked lines,
there is a complete discussion in Peterson et al.~2004).
The line dispersion, which can be calculated without making any assumption about the line profile, can be defined by:
\begin{equation}\label{equ:2moment}
    \sigma^{2}_{line}(\lambda)=<\lambda^{2}> - \lambda_{0}^{2}
\end{equation}
where $<\lambda^2>$ and $\lambda_{0}$ are the $2^{nd}$ \& $1^{st}$
moments of the emission line profile $P(\lambda)$:
\begin{eqnarray}
  <\lambda^{2}>=\frac{\int\lambda^{2}P(\lambda)d\lambda}{\int P(\lambda)d\lambda} \\
  \lambda_{0}=\frac{\int\lambda P(\lambda)d\lambda}{\int
  P(\lambda)d\lambda}.
\end{eqnarray}
The second moment is more sensitive to blending with other lines
or extended wings than the FWHM, but less sensitive to the
presence of even strong narrow-line components. The FWHM
versus time lag plots for emission lines show dramatic scatter
(especially for \Hbeta) when using the mean spectra (which are more similar
to single epoch spectra than rms spectra are), while the
line dispersion, $\sigma_{line}$, versus time lag plots show less scatter (Peterson et
al.~2004).

On the other hand, studies of RM samples have shown that AGNs with
line dispersion, $\sigma_{line}$, lower than 2000~\kmps\ have
$\textrm{FWHM}/\sigma_{line}<2.35$ while AGNs with
$\sigma_{line}>2000$~\kmps\ have $\textrm{FWHM}/\sigma_{line}>2.35$
(Sulentic et al.~2000; Collin et al.~2006; Peterson et al.~2007).
In particular, Peterson et al.~(2007)
report that the observed value of $\textrm{FWHM}/\sigma_{line}$
for a sample of RM AGN ranges between $\sim0.71$ and $\sim3.45$
with an average value of $\sim2.0$. This result clearly indicates the
non-Gaussianity of line profiles for quasar emission lines.
Thus using the FWHM as a line width indicator may overestimate the BH masses for
the broadest emission lines and underestimate the BH masses for the narrowest
emission lines as compared to BH mass estimates using the line dispersion.

The true line dispersion can be determined through
considering the instrumental resolution and writing the observed
line width in terms of the intrinsic line dispersion,
$\sigma_{intr}$ and the spectrograph's instrumental line profile,
$\sigma_{inst}$:
\begin{equation}\label{equ:sig_true}
\sigma^{2}_{line} = \sigma^{2}_{intr} +
(1+z)^{-2}\sigma^{2}_{inst}
\end{equation}
where the instrumental dispersion is observed-frame while the
other two are rest-frame. The instrumental resolution of an SDSS
spectrum is a function of plate, fiber, and wavelength.  The
wavelength dispersion in pixels as a function of wavelength for
each fiber on a plate is stored in HDU \#4 of the {\em spPlate}
FITS files.
For our objects, we need to know the instrumental dispersion at the observed
wavelength of \MgII.  For each quasar, the observed wavelength of
\MgII\ is calculated as $\lambda_o=2800.26(1+z)$. Using the {\em
spPlate} file, the corresponding pixel in the quasar's SDSS
spectrum is located and the wavelength dispersion $d$ (in units of
pixels) at that pixel is extracted. The pixel scale is $10^{-4}
\ln (10) \lambda_o$ \AA/pixel, so the instrumental dispersion in
\AA\ is given by $2.302585 \times 10^{-4} \lambda_o d$.

%=======================================================================================================
\subsection{Fitting Method}\label{sec:FM}

The rest-frame ultraviolet spectrum of an AGN can be approximated as a power-law continuum plus broad and overlapping \FeII\ emission and
broad emission lines from other ions such as \MgII. To measure the \MgII\ line, we must remove the contaminating \FeII\ emission from AGN spectra. We subtract a best-fit \FeII\ template and power law continuum estimate, which together form what is called a pseudo-continuum.

To match the \FeII\ lines to the AGN spectra, we build an \FeII\ model by convolving the \FeII\ template generated from the nearby AGN I~Zw~1 by Vestergaard \& Wilkes~(2001) with a Gaussian.
The AGN spectra have lower resolution than the \FeII\ template so we rebin the \FeII\ template to match the AGN spectra in their rest frames.
Vestergaard \& Wilkes~(2001) has been the standard empirical \FeII\ template
since its publication.  Tsuzuki et al.~(2006) present a new template,
also derived from I~Zw~1, which unlike that of Vestergaard \& Wilkes~(2001)
shows some \FeII\ emission at wavelengths 2780-2920~\AA\ where
confusion with \MgII\ is an issue in constructing a template.  While some
such \FeII\ emission is certainly present, Tsuzuki et al.~(2006) derive
their \FeII\ template by {\em assuming} Gaussian line profiles for \MgII\
in I~Zw~1.  What they interpret as \FeII\ emission at 2790~\AA\ could
instead be a weak blue wing of \MgII\ emission.
Furthermore, neither the theoretical \FeII\ spectra considered by
Tsuzuki et al.~(2006) nor those presented by Bruhweiler \& Verner~(2008)
show strong evidence for emission at wavelengths corresponding to features
at 2780-2920~\AA\ in the Tsuzuki et al.~(2006) template.
More work on \FeII\ templates is clearly needed, but meanwhile we will use
the Vestergaard \& Wilkes~(2001) template.

The normalization and slope of the power law continuum
are initially estimated using two normalization windows, $2238-2248$ {\AA} and $3014-3027$ {\AA}, which are the least contaminated from \FeII\ or other lines (Vestegaard and Wilkes~2001).

To allow improvement in the fitting quality we scale the \FeII\ template independently above and below $2800${\AA}.
This additional amplitude will be accepted if the new $\chi^{2}$ is lower than the 95\% confidence lower limit on the old $\chi^2$ with only 4 free parameters. The best pseudo-continuum is then subtracted from the AGN spectra.

We vary the amplitude and slope of the power law continuum,
the amplitude of the \FeII\ template above and below $2800${\AA},
and the width of the Gaussian convolved with the \FeII\ template
to find the best fit.
Using a nonlinear Levenberg-Marquardt least-squares fitting method, we calculate the minimum $\chi^2$ value for the three fitting windows $2192-2400$ {\AA}, $2445-2724$ {\AA}, and $2867-3027$ {\AA} chosen to avoid the \NeIV\ $\lambda$2423 {\AA} and \MgII\ $\lambda$2798 {\AA} emission lines during the \FeII\ and continuum estimation.

%=======================================================================================================
\section{Determining the Mass Scaling Relationship} \label{sec:Scaling relationship}
To determine a mass scaling relationship which uses $\sigma_{MgII}$ and $\lambda L_{3000}$,
we first need to investigate the existence of a relationship between $\lambda L_{3000}$ and $R_{BLR}$. If that relationship exists then we can scale our mass relationship to an existing sample of BH masses to determine the geometrical factor.

%--------
The most direct measurements of the central
black hole masses of powerful AGN come from the reverberation
mapping studies of $17$ Seyferts and $17$ PG quasars by Wandel,
Peterson \& Malkan~(1999) and Kaspi et al.~(2000), respectively. These studies are
based on the time lag measurement between the variation
of the AGN continuum and the broad emission line \Hbeta\ of local AGN (low redshift objects with $z<0.4$).
We thus use this sample as our reference for the scaling process.

We select $21$ objects from this sample (Table 1 of McLure \& Jarvis~2002)
for which the \MgII\ \textrm{FWHM} has been calculated. We also have
\textrm{FWHM} and \Hbeta\ line dispersion measurements for the same objects
from Peterson et al.~(2004).

This list of AGN consists of 5 high resolution spectra from the Hubble Space Telescope (HST) Faint Object Spectrograph (FOS)
and 16 low resolution spectra from the International Ultraviolet Explorer (IUE)
Long-Wavelength Prime (LWP) spectrograph (see Table \ref{tab:tbl-1}). For most of them there is more than one observation. In order to increase the signal-to-noise ratio and sometimes to increase the wavelength coverage, we combine different observations of each object by using
a weighted average method.  We adopt as the time-lag to the Mg\,{\sc ii} BLR the
weighted average time-lag estimated using H$\alpha$, H$\beta$ and H$\gamma$,
or whichever subset of those three are available, from Peterson et al. (2004).
While these lines sometimes have different timelags, in this case
the small number of H$\alpha$ and H$\gamma$ measurements available means
that the resulting correlation is not significantly different if only the
H$\beta$ timelags are used.

By using our fitting procedure, we find the best-fit broadened \FeII\ template and the best power law continuum for each spectrum in Table \ref{tab:tbl-1} to create the corresponding pseudo-continuum. We subtract this pseudo-continuum from the spectrum so that we measure the line dispersion of the \MgII\ from the isolated emission line. The line dispersion for \MgII, $\lambda L_{3000}$, and the virial mass estimate can be seen in Table \ref{tab:tbl-1} column 3, 4, and 8 respectively.

%--------
In a second approach, we use PCA-reconstructed spectra instead of raw spectra
(see \S\,\ref{sec:RQS-QE} for details).
However, using reconstructed spectra with high SNR does not improve the $\tau-\lambda L_{3000}$ correlation significantly.

We estimate the slope and intercept of the linear correlation between the logarithms of $\tau_{BLR}$ and $\ell\equiv \lambda L_{3000}/10^{44}$~erg\,s$^{-1}$ using several fitting techniques.
However, we eliminate some of the objects from our list
based on a large relative error of the average time-lag,
$\Delta\tau_{average}/\tau_{average} > 31.7\%$.
These objects are %include
3C390.3, PG$0844+349$, PG$1229+204$, NGC4051 and 3C120.

\begin{figure}
  \centering
  \includegraphics[scale=0.8,angle=0]{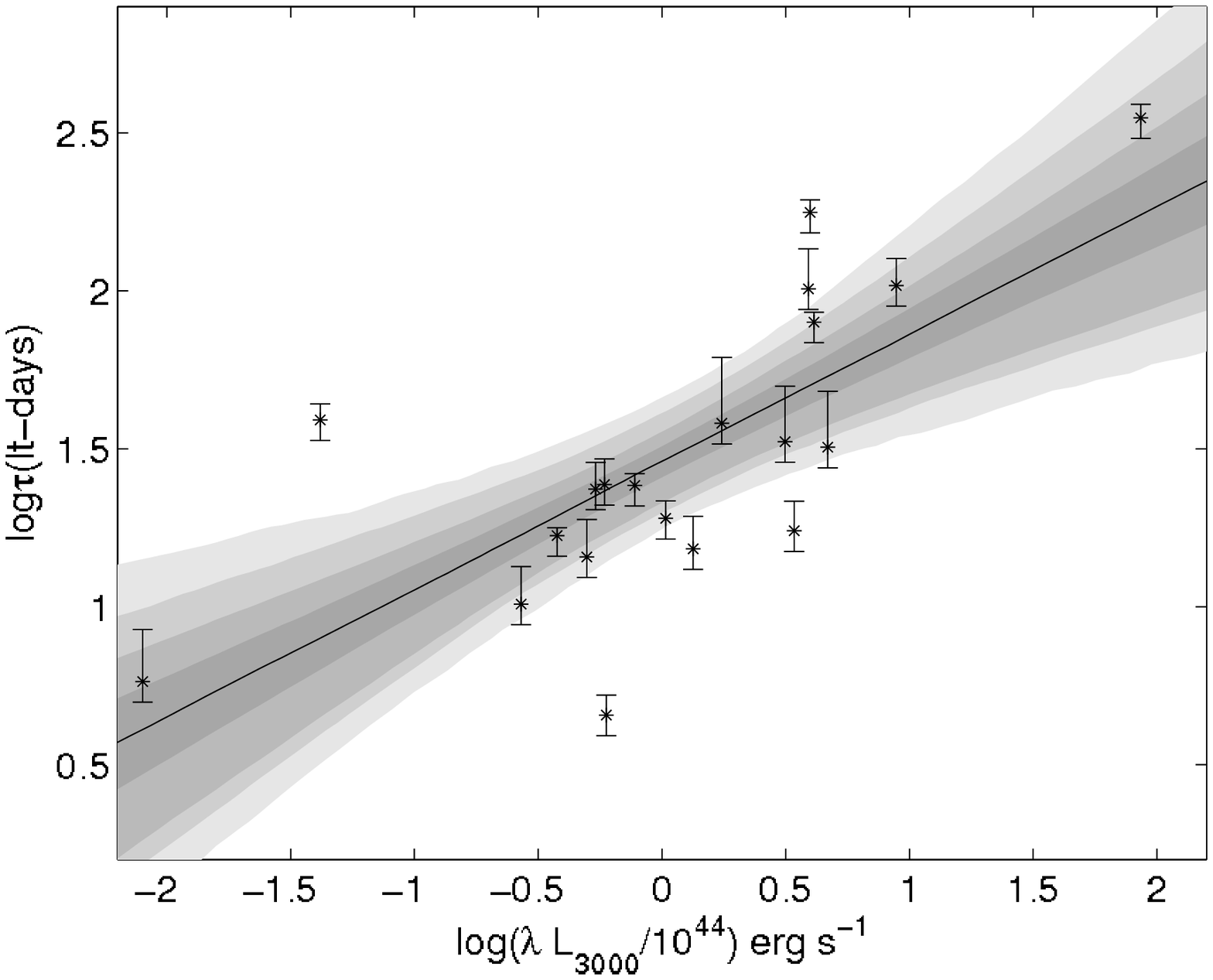}
   \caption{The best linear fit to the reference sample in Table \ref{tab:tbl-1} using Markov Chain Monte-Carlo Simulation (MCMC; Haario et al.~2006) method with $33\%$ intrinsic scatter. The gray areas correspond to: 50\%, 90\%, 95\% and 99\% posterior regions.}\label{fig:RL-correlation}
\end{figure}

Both the FITEXY (Tremaine et al.~2002; Press et al.~1992)
and BCES methods (Akritas et al.~1996) gave a slope consistent with 0.5 within $\approx1\sigma$, with 20\% uncertainty in the best case (see Table \ref{tab:fitting-results}).
Given those results,
and because a slope of 0.5 matches the theoretical slope suggested by Netzer
et al. (1993) for a fixed ionization parameter at the innermost radius of
the BLR and the observational slopes found for H$\beta$ (Bentz et al. 2006)
and C\,{\sc iv} (Kaspi et al. 2007), we chose to adopt a fixed slope of 0.5
and recalculate the intercept from the data.
In Figure \ref{fig:RL-correlation} we show the $\tau_{BLR}-\ell$ correlation calculated by MCMC\footnote{Markov Chain Monte-Carlo Simulation} along with the confidence regions for the correlation (Haario et al.~2006). The correlation that we adopt is:
\begin{eqnarray}\label{equ:logR-LL}
 \log(\tau_{BLR})= 0.5 ~\log(\ell)+ (1.469\pm0.090),\\ \nonumber\\
\tau_{BLR}(\textmd{days})=(29.44^{+6.78}_{-5.51})~ \ell^{0.5},
  \nonumber\\
  \ell=[\lambda~L_{3000}/10^{44} \textmd{\ergps}].
\end{eqnarray}
for $\chi^{2}_{\nu}\approx1$ with an intrinsic scatter of
$\sim35\%$.
Some of this intrinsic scatter arises because the luminosity measurements
are not necessarily co-temporal with the time-lag measurements.  However,
quasar variability on timescales of 100 or more rest-frame days can be roughly
described by a Gaussian with $\sigma_m=0.2$ magnitudes (Figure 12 of
Vanden Berk et al. 2004).
The corresponding additional scatter in the $\tau_{BLR}$-$\ell$
relationship should therefore be roughly described by a Gaussian
with $\sigma = 0.04$ dex, or about 10\%.

%=======================================================================================================
McLure and Jarvis (2002) estimated a correlation for $R_{BLR}$ with luminosity at
3000{\AA} by using the reverberation mapping data of 34 quasars
from Wandel et al. (1999) and Kaspi et al. (2000) and found
 $R_{BLR}=(25.2\pm3.0)\ell^{0.47\pm0.05}$
where $R_{BLR}$ is in units of light days.
Both of our correlations are well defined only for quasars with $0.1<z<0.5$ within the luminosity
range $0.001<\ell<100$. However, we assume that our
correlation will hold for objects with luminosity and redshift beyond those limits
(as tentatively found for C\,{\sc iv} by Kaspi et al. 2007,
but see appendix A1 of McLure \& Dunlop 2004 for a contrary result).
%=======================================================================================================

We can determine the average geometrical factor, $\bar{f}$,
in Equation 1 through comparing our virial mass estimates with
the reverberation mass estimates by Peterson et al.~(2004), which are calibrated to the
\Msigma\ correlation as described in Onken et al.~(2004).\footnote{Note that $\bar{f}$ is not exactly the same as $f$, the geometrical factor used to scale reverberation-mapping virial products to the $M_{BH}-\sigma_*$ relationship.  We use $\bar{f}$ to scale single-epoch virial products to calibrated RM masses which already incorporate $f$.}
We assume a linear fit with
a fixed zero intercept for the virial product, $VM_{BH}=R_{BLR}\sigma_{intr}^2/G$, versus reverberation black hole mass, $RM_{BH}$.
(A fit where the intercept is allowed to vary yields an intercept consistent with zero at $\ll 1\sigma$).
The reduced chi squared, $\chi^{2}_{\nu}$, is calculated from:
\begin{equation}\label{equ:chi2nu-f}
  \chi^{2}_{\nu}=\frac{1}{N-1}\sum_{i=1}^{N}\frac{[(VM_{BH})_{i}\bar{f} - (RM_{BH})_{i}]^{2}}{\sigma_{VM}^{2}+\sigma_{RM_{BH}}^{2}+\epsilon_{0}^{2}}
\end{equation}
where the geometrical factor - which is the slope of the linear fit -
is $\bar{f}= 5.3\pm0.6$ with $\chi_{\nu}^{2}\simeq 1.92$ ($N=15$)
considering an intrinsic scatter of $\epsilon_{0}\simeq35\%$ for the estimated masses
from both methods.
The relationship between mass, luminosity and linewidth is:
\begin{eqnarray}\label{equ:M-LL-sigma}
   \textrm{\mbh}/\textrm{\msol}=5.75\bar{f} \ell^{0.5}\sigma_{intr}^{2}
   %\nonumber\\
   \pm\sigma_{\textrm{\mbh}/\textrm{\msol}}(\textrm{statistical})
   \nonumber\\
   \pm24\% (\textrm{systematic})
   \nonumber\\
   \pm35\% (\textrm{intrinsic~scatter})
\end{eqnarray}
where the constant 5.75 is the conversion factor for using BH masses in solar units when $\ell=[\lambda L_{3000}/10^{44}]$ is in units of \ergps, and the intrinsic line dispersion of the \MgII\ emission line, $\sigma_{intr}$, is in units of \kmps. The statistical error of the black hole mass estimates, $\sigma_{\textrm{\mbh}/\textrm{\msol}}$, can be calculated from:
\begin{eqnarray}\label{equ:sigmaMBH}
\sigma_{\textrm{\mbh}/\textrm{\msol}}=2.875[\frac{\sigma^{2}_{intr}\bar{f}^2}{\ell}(
\sigma_{\ell}^{2}\sigma_{intr}^{2} %\nonumber\\
+16\sigma_{\sigma_{intr}}^{2}\ell^{2})]^{0.5}.
\end{eqnarray}
where $\sigma_{\sigma_{intr}}$ and $\sigma_{\ell}$ are estimated
errors from our fitting process. The scatter of 35\% is equivalent to 0.15 dex.

The systematic error of 24\% (equivalent to 0.10 dex) comes from propagating the relative errors of the geometrical factor $\bar{f}$ and
of the coefficient in the $\tau_{BLR}-\ell$ relation (Equation \ref{equ:logR-LL}). It does not include any uncertainty in the exponent of the $\tau_{BLR}-\ell$ relation.

%=======================================================================================================
\section{The Sample}\label{sec:The_Sample}

In this study we use using spectra from the
SDSS Data Release Three (DR3; Abazajian et al. 2005).
The SDSS wavelength coverage sets the upper and lower limits for our sample selection.
A spectrum needs to have the rest-frame wavelength of 2192-3027~{\AA}
within the SDSS range, since we fit that region around the \MgII\ emission line (2798 {\AA}) with a power law continuum and \FeII\ emission template.

\begin{figure}
  \centering
 \includegraphics[scale=0.6,angle=0]{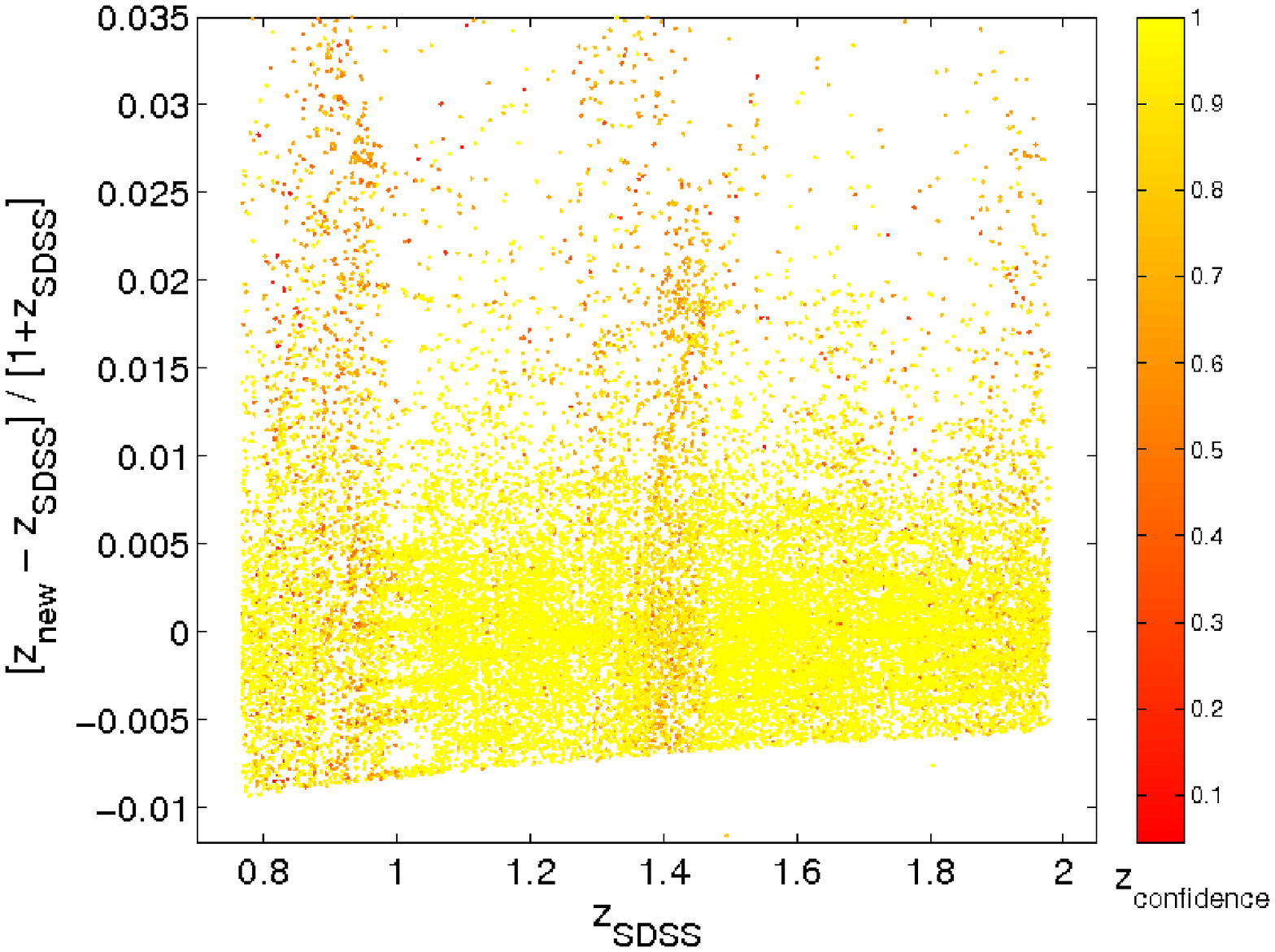}
  \includegraphics[scale=0.6,angle=0]{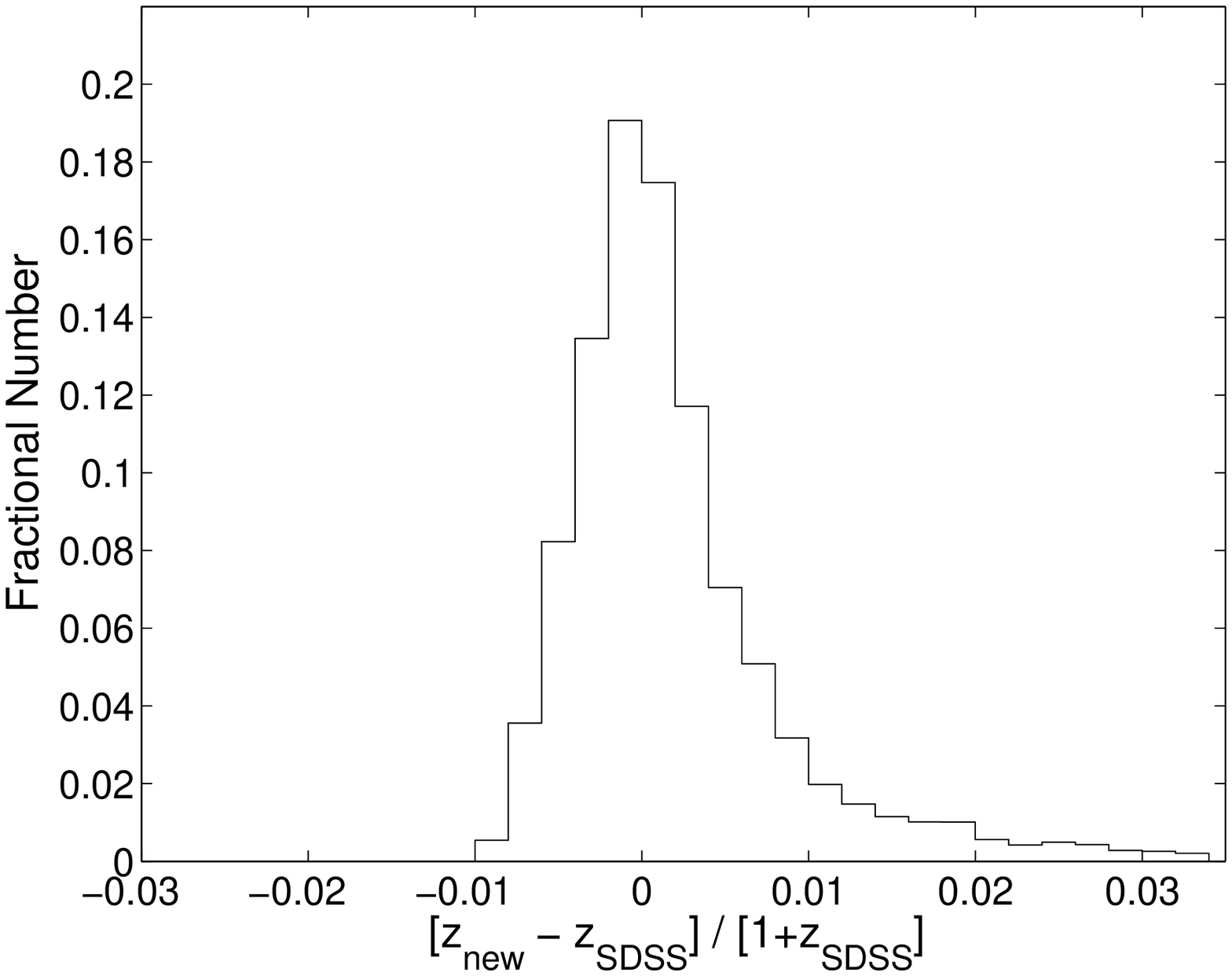}
  \caption{The top panel shows the relative change in quasar redshift after cross correlating the median quasar composite spectrum (Vanden Berk et al.~2001) with SDSS quasar spectra versus the SDSS redshift estimates described in \S 4.10.2.1 of Stoughton et al.~(2002). The colormap is based on the estimated redshift confidence from the SDSS catalogue (Stoughton et al.~2002). The bottom panel shows the distribution of the relative change in quasar redshift. See the online version for color Figures.
  }\label{fig:hist_redshift_oldnew}
\end{figure}

The above condition defines our acceptable redshift range, which is $0.7336<z_{SDSS}<2.0393$.
However, the redshift given in the DR3 quasar catalog (Schneider et al. 2005),
as described in \S 4.10.2.1 of Stoughton et al.~(2002), is not always as accurate as we need to place the \MgII\ emission line sufficiently close to $2800$~{\AA} in the adopted rest frame.
Thus, we cross correlate the median quasar composite spectrum (Vanden Berk et al. 2001) with SDSS DR3 spectra
(using the IRAF\footnote{The Image Reduction and Analysis Facility, http://iraf.noao.edu/}
task "xcsao")
to redefine the object's redshift. These corrected
redshift will be reliable only when relative change of the redshifts
is smaller than $2\%$. Consequently the new redshift range will come
from $|(z_{new}-z_{SDSS})|/(1+z_{SDSS})\simeq 2\%$. The top panel in Figure \ref{fig:hist_redshift_oldnew} shows the relative change in redshifts versus SDSS DR3 redshift, color-coded with the redshift confidence from the SDSS DR3 (Stoughton et al.~2002). Indeed, redshifts with lower confidence in SDSS DR3 tend to have larger redshift corrections. The bottom panel in Figure \ref{fig:hist_redshift_oldnew} shows the distribution of both redshift estimates.

These new redshift limits, $0.6996<z_{new}<2.1013$, however, include those
objects where after redshift correction they have been shifted to the
acceptable wavelength range (or maybe shifted out of the range), where
the acceptable range is still defined by the normalization region.

The luminosities of our calibration objects come from total
magnitudes, but SDSS DR3 spectra are scaled only to the flux in a 3\arcsec\ diameter aperture.
The average correction factor we apply to the flux measured from SDSS DR3
spectra to produce the total flux of all our objects is $-$0.35 magnitudes,
which is appropriate for unresolved objects in the SDSS DR3
(see section 4.1.1 of Adelman-McCarthy et al. 2008).
While this is an average correction, it is appropriate for quasars at all
redshifts and luminosities in our sample ($z>0.7$ and $M_i<-24$),
as the fraction of SDSS quasars with extended morphologies at those
luminosities is at most 2\% (Figure 10 of Vanden Berk et al. 2006).
%However, No such correction is needed for SDSS spectra released later than DR3 because their flux levels are
%scaled to PSF magnitudes, which represents the best estimate of the total flux
%for unresolved objects in the SDSS (Adelman-McCarthy et al. 2008).

Using our new quasar redshifts,
the percentage of the successful fits in DR3 sample is about $79\%$ of the initially selected sample.
To increase the fraction of acceptable fits and to reduce the noise we later use the eigenspectra
method to reconstruct the quasar spectra (see section \ref{sec:RQS-QE}).
This increases the percentage of acceptable fits to $94\%$.
The remaining $6\%$ have problems such as a lack of data exactly at the
\MgII\ region, or a very broad absorption line very close to
\MgII\ or the normalization regions.
In the end, our DR3 sample consists of 93.3\% of all quasars with
$0.7336 < z_{SDSS} < 2.0393$ (27602 quasars out of 29582).

Note that the objects in our catalogue are not homogeneously selected.
However, of the 9325 quasars in redshift range in the homogeneous
subsample of Richards et al. (2006), our DR3 sample contains 8814 (94.5\%).
Quasars in that complete subsample are flagged in the DR3 mass catalogue,
and only such quasars should be used for studies that require
a volume-limited, homogeneously selected sample.
However, even in such studies the effects of Malmquist bias must
be taken into account (see \S \ref{sec:Malmquist Bias}).

%=======================================================================================================
\section{Reconstructing quasar spectra from quasar eigenspectra}\label{sec:RQS-QE}

Principal Component Analysis (PCA), sometimes called the Karhunen-Lo\`{e}ve
transformation,
is an orthogonal linear transformation of data to a new coordinate system.
As a result of this transformation the greatest variance by any projection of the data forms the first coordinate (called the first principal component or first eigenspectrum), the second greatest variance forms the second coordinate, and so on. Thus the great advantages of using PCA is that a number of possibly correlated variables transform into a smaller number of orthogonal variables called principal components.

Yip et al.~(2004) have applied this transformation to a sample of 16707 quasar spectra from the SDSS DR1 to generate such principal components. They have performed the transformation on subsamples with different redshift ($z$) and absolute magnitude ($M_i$) in the $i$ filter to generate eigenspectra for each ($M_i,z$) bin.
Using this technique they separate any possible luminosity effect on the spectra from any evolutionary effects.
In each ($M_i,z$) bin, the flux values from all quasars as a function of rest-frame wavelength are the data points on which the PCA is run. Therefore, PCA generates eigenspectra which span the same
rest-frame wavelength range as the input spectra.
Yip et al.~(2004) show that the first four eigenspectra in each ($M_i,z$) bin account for 82\% of the total sample variance.
These four eigenspectra represent the mean spectrum, a host-galaxy component, UV-optical continuum slope variations, and the correlations of Balmer emission lines respectively.
We used all 50 eigenspectra calculated by Yip et al.~(2004) in the appropriate ($M_i$,$z$) bin to provide the most accurate reconstruction of our spectra possible.  We considered using only the number of eigenspectra needed to reach $\chi_\nu^2=1$ for each object, but such an approach is not possible for all objects with the Yip et al.~(2004) eigenspectra because Figure \ref{fig:chi2_PCA_estimate} shows that most of our objects would require more than 50 eigenspectra to reach $\chi_\nu^2=1$.

\begin{figure}
  \centering
  \includegraphics[scale=0.9,angle=0]{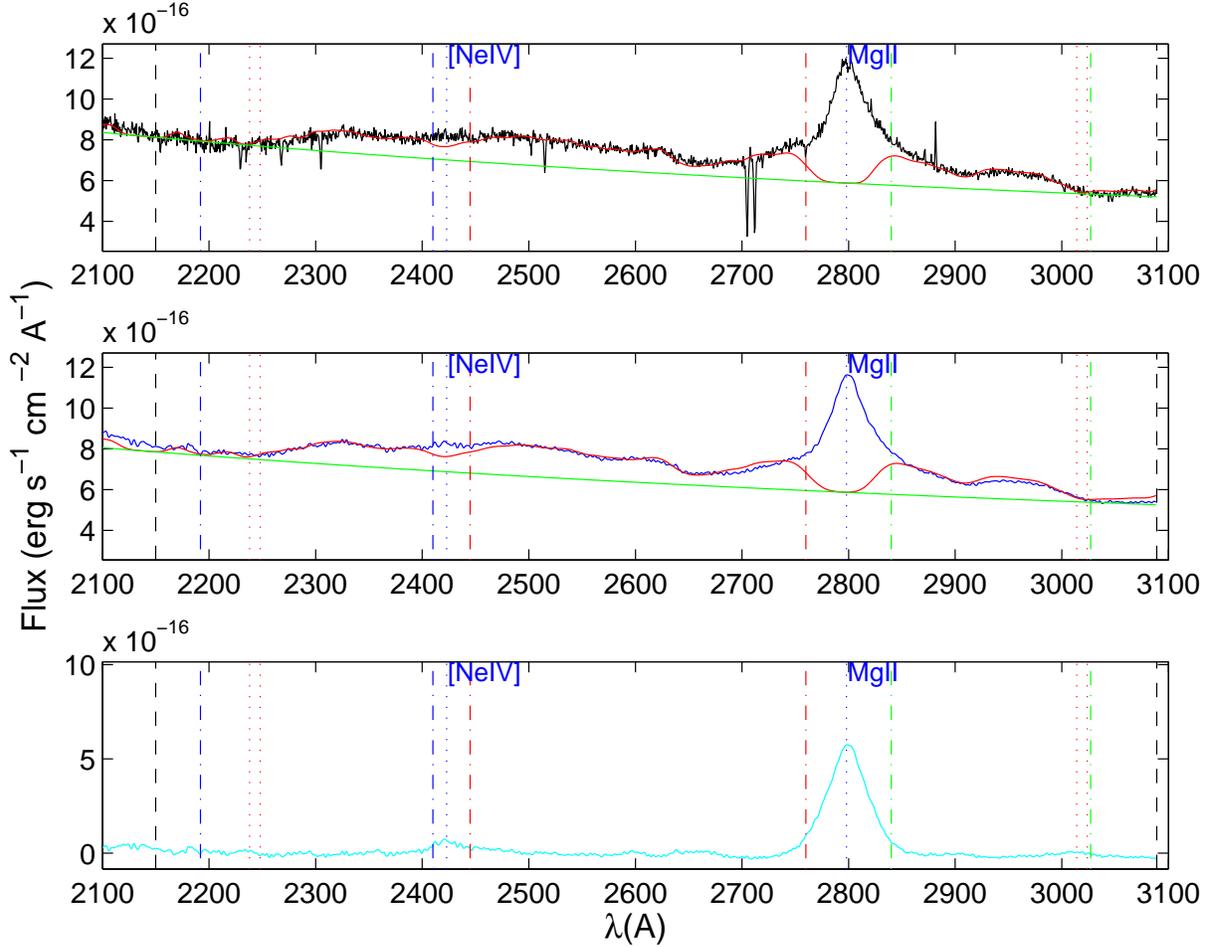}
   \caption{A sample SDSS quasar spectrum, before and after reconstruction using PCA, and our best pseudo-continuum estimates for it. The top panel shows the observed SDSS spectrum (black) and continuum (green) and the best pseudo-continuum fit (red). The middle panel shows the PCA reconstructed spectrum (blue) and continuum (green) and the best pseudo-continuum fit (red). The bottom panel shows the reconstructed spectrum in cyan after subtracting the best pseudo-continuum. The dotted red vertical lines show the normalization windows where we initially estimate the normalization of the power law, $2238-2248$\,\AA\ and $3014-3027$\,\AA. The dotted blue lines show two emission lines \MgII\ and \NeIV. The dot-dashed lines determine the ranges within which we calculate the $\chi^2$ of the pseudo-continuum fit. The black dashed lines are the borders of the fitting region. Narrow absorption lines and noise spikes within the \FeII\ fitting region have been removed through PCA reconstruction. See the online version for color Figures.}\label{fig:SDSS_sample}
\end{figure}

We have used the 50 most significant eigenspectra from Yip et al.~(2004) (the luminosity, redshift dependent eigenspectra) to reproduce the spectrum of each quasar. The reproduced spectra have the same continua as the original spectra, and  the most
significant features like emission lines are the same. However, the reconstructed spectra have less noise as compared to the original spectra.
Any narrow absorption lines are removed through this method, and narrow
regions of missing data are effectively reconstituted by the best-fit
reconstruction found for the rest of the spectrum.
In Figure \ref{fig:SDSS_sample} we show an example of an SDSS quasar spectrum before and after PCA reconstruction.
The top panel shows the raw spectrum and our best pseudo-continuum estimate for it, and the middle panel the same for the reconstructed spectrum. The bottom panel shows the decontaminated spectrum which we use to measure the \MgII\ line dispersion.

In Figure \ref{fig:chi2_PCA_estimate} we show the distribution of the reduced $\chi^2$ calculated for DR3 catalogue by comparing reconstructed spectra and raw spectra for two cases: first, using a window around \MgII\ between $2700${\AA} and $2900${\AA} and second, using the entire fitting region of 2192-3027{\AA}. For both cases, the peak is located near unity, indicating that the PCA reconstructions are statistically
acceptable fits to the original spectra.

\section{Radiation Pressure Correction (RPC) on Scaling Relationship}\label{sec:Radiation Pressure Correction}
The virial theorem can be used for a dynamically relaxed gravitationally bound system. However, very luminous sources can violate this assumption.
In such sources, the gravitational force of a BH can be reduced by the effect of the radiation pressure (Marconi et al. 2008). If BLR clouds are attracted by a smaller effective force of gravity, their line dispersion decreases, resulting in mistakenly underestimated virial BH mass estimates. Therefore, the scaling relationships may therefore need to be recast to include the effect of radiation pressure.

We have applied the Marconi et al.~(2008) suggestion to correct the scaling relationship for the \MgII\ emission line such that $\textrm{\mbh}/\textrm{\msol}=5.75\bar{f}_{a} \ell^{0.5}\sigma_{intr}^{2}+ \bar{g} \ell$ where $\ell=[\lambda L_{3000}/10^{44}]$\ergps and $\sigma_{intr}$ is the line dispersion of the \MgII\ line in units of \kmps.
%----
We use the same sample as in \S\,\ref{sec:Scaling relationship}
to determine the adjusted calibration factors, $\bar{f}_{a}$ and $\bar{g}$,
through comparison of our virial mass estimates adjusted for radiation pressure with the black hole mass estimates $M_{BH}^*$ made directly from the black hole mass - bulge velocity dispersion relationship for quiescent galaxies (Equation 1 in Tremaine et al.~2002). We assume that the relationship holds for AGNs and their host galaxies too. The velocity dispersions $\sigma_{*}$ are extracted for 11 objects from Onken et al.~(2004) and for two objects, PG1229+204 and PG2130+099, from Dasyra et al.~(2007). We assume a non-linear fit with a variable intrinsic scatter for the Adjusted Virial Mass ($AVM_{BH}$) versus $M_{BH}^{*}$. The reduced chi squared, $\chi^{2}_{\nu}$, is calculated from:
\begin{equation}\label{equ:chi2nu-fg}
  \chi^{2}_{\nu}=\frac{1}{N-3}\sum_{i=1}^{N}\frac{[(AVM_{BH})_{i} - (M_{BH}^{*})_{i}]^{2}}{\sigma_{VM}^{2}+\sigma_{M_{BH}^{*}}^{2}+\epsilon_{0}^{2}}
\end{equation}
Where the adjusted calibration factors are $\bar{f}_{a}= 2.4\pm1.5$ and $\log\bar{g}=6.9\pm0.5$ with $\chi_{\nu}^{2}\simeq 1.00$ ($N=12$; after excluding 3C390.3) considering an intrinsic scatter of $\epsilon_{0}\simeq15\%$. The adjusted mass scaling relation using \MgII\ emission line is:
\begin{eqnarray}\label{equ:RPC}
   \textrm{\mbh}/\textrm{\msol}=5.75 \bar{f}_{a} \ell^{0.5}\sigma_{intr}^{2} + \bar{g} \ell
   \nonumber\\
   \pm\sigma_{\textrm{\mbh}/\textrm{\msol}}(\textrm{statistical})
   \nonumber\\
   \pm15\%(\textrm{intrinsic~scatter}).
\end{eqnarray}
The statistical error of the black hole mass estimates, $\sigma_{\textrm{\mbh}/\textrm{\msol}}$, can be calculated from:
\begin{eqnarray}\label{equ:err_RPC}
 \sigma_{\textrm{\mbh}/\textrm{\msol}}=[\sigma_{\ell}^{2}(2.875\frac{\bar{f}_{a}\sigma_{intr}^{2}}{\ell^{0.5}}+\bar{g})^2
 +132.2\sigma_{\sigma_{intr}}^{2}\sigma_{intr}^{2}{\bar{f}_a}^2\ell)]^{0.5}
\end{eqnarray}
where $\sigma_{\sigma_{intr}}$ and $\sigma_{\ell}$ are estimated errors from our fitting process. The scatter of 15\% corresponds to less than 0.07 dex of intrinsic scatter. We do not quote a systematic error (which comes from the systematic errors on $\bar{f}_{a}$ and $\bar{g}$) because its magnitude depends on the relative values of the two terms in Equation \ref{equ:RPC}, which will be different for objects of different luminosities and linewidths.

When the estimated line dispersion of \MgII\ is very small then $\lim_{\sigma \rightarrow 0} \textrm{\mbh}/\textrm{\msol}=\bar{g}\ell$. Using our luminosity limits we have a lower limit on black hole mass estimates for the adjusted method of $\log(\frac{\textrm{\mbh}}{\textrm{\msol}})\gtrsim 6.9$ for $\lambda L_{3000}>10^{44}$ \ergps. Whether this is a real constraint remains an open question.

\section{Noise Simulation}\label{sec:Noise simulation}

\begin{figure}
\centering
  \includegraphics[scale=0.8,angle=0]{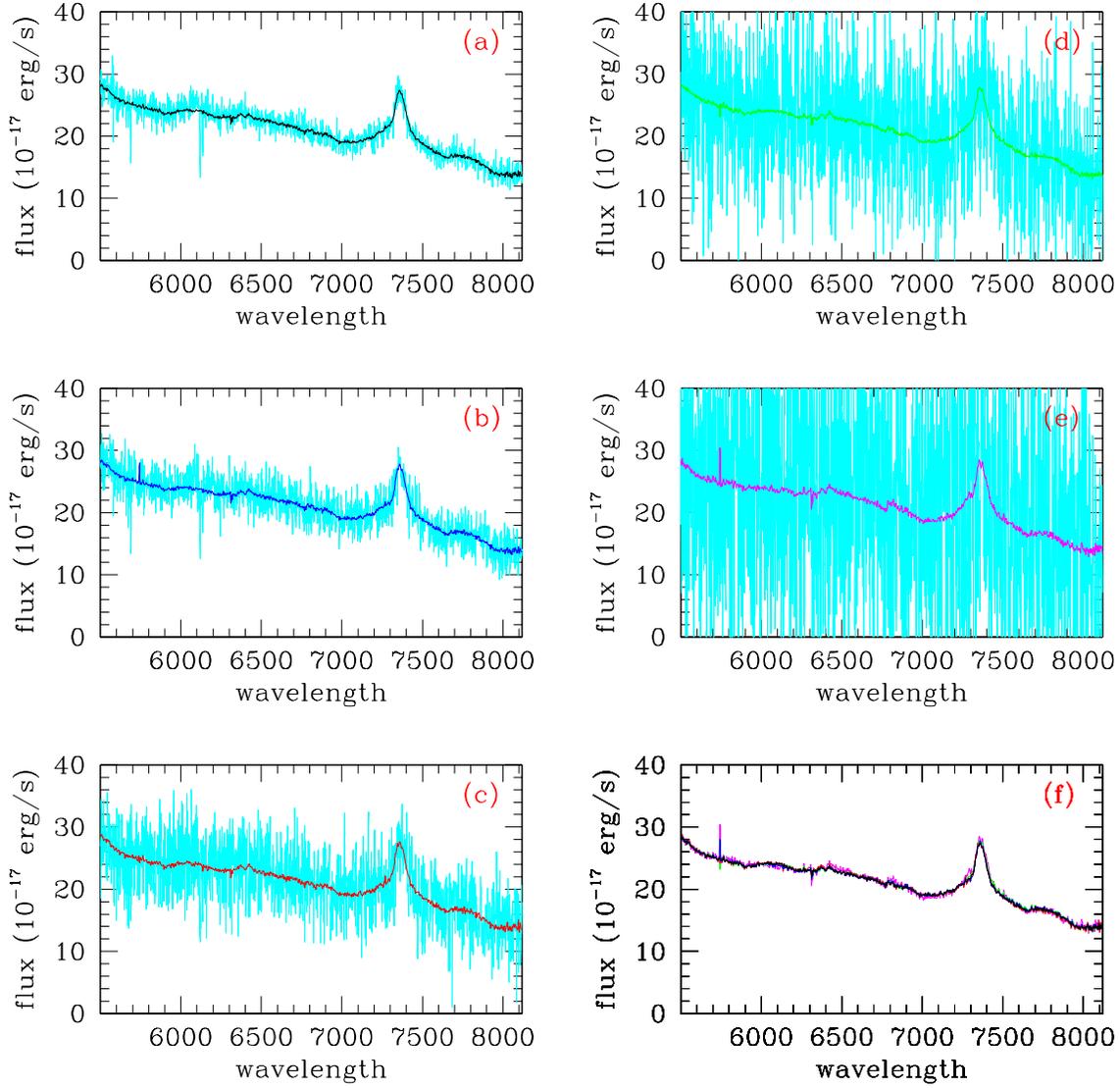}\\
  \caption{A typical quasar spectrum with noise added to produce spectra with different SNR levels (cyan). The spectrum reconstructed in each case using the 50 most significant PCA eigenspectra from Yip et al.~(2004) is overplotted. (a) Reconstructed spectrum shown in black generated from spectrum with original SNR in cyan. (b) Reconstructed spectrum in blue generated from spectrum with $1/2$ original SNR. (c) Reconstructed spectrum in red generated from spectrum with $1/4$ original SNR. (d) Reconstructed spectrum in green generated from spectrum with $1/8$ original SNR. (e) Reconstructed spectrum in magenta from spectrum with $1/16$ original SNR. (f) over-plotting the 5 reconstructed spectra in panel (a) to (e). See the online version for color Figures.}\label{fig:noise_added_spectra}
\end{figure}

\begin{figure}
\centering
   \includegraphics[scale=0.8,angle=0]{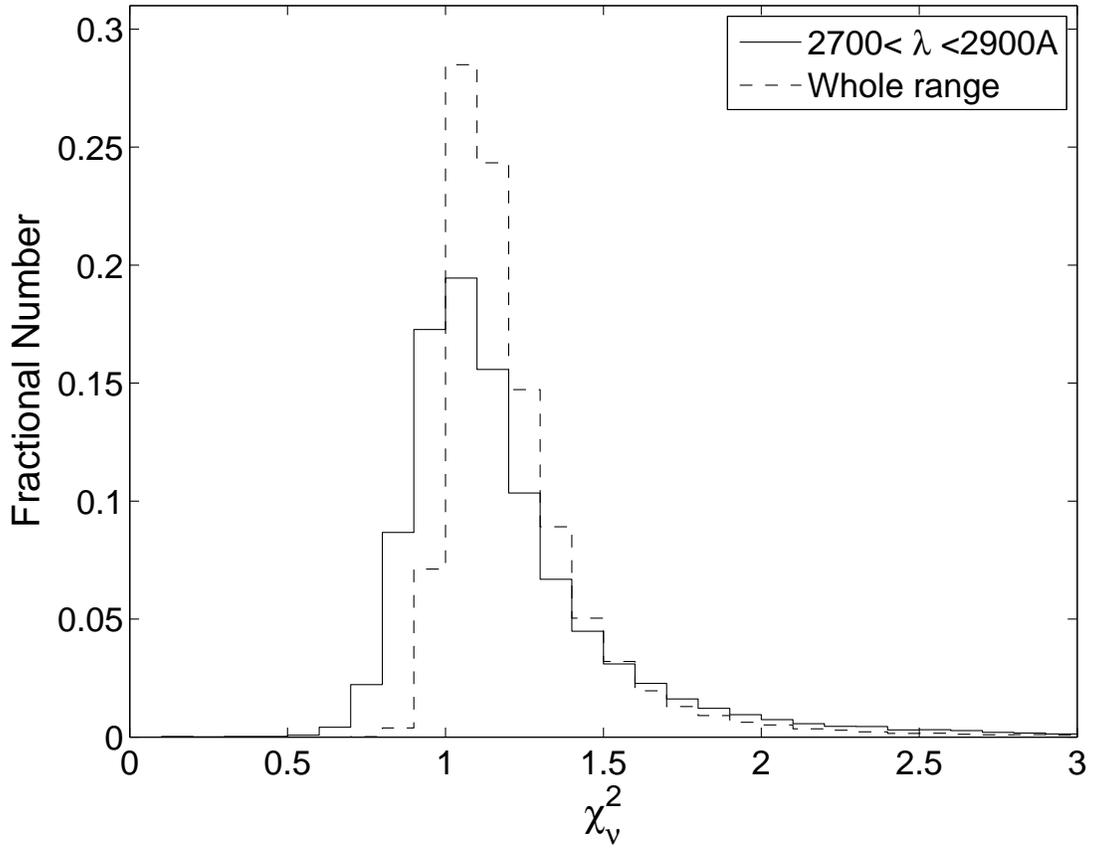}\\
  \caption{The distribution of the $\chi^2_{\nu}$ comparing the reconstructed spectra using PCA technique with raw spectra for two cases: solid curve compares within the window $2700$~{\AA}$< \lambda < 2900$~{\AA}, dash-line curve compares the whole spectrum. See the online version for color Figures.}\label{fig:chi2_PCA_estimate}
\end{figure}

We now evaluate the statistical errors on BH mass estimates from reconstructed spectra due to the presence of different noise levels in quasar spectra before reconstruction. We select 100 quasars with the highest signal to noise ratio (SNR), from 20 to 50. Then we add randomly generated Gaussian noise to these spectra to decrease the SNR to 1/2, 1/4, 1/8, and 1/16 of the original SNR (see Figure \ref{fig:noise_added_spectra}).
We repeat this process for each new SNR 6 more times to have at least 25 realizations for every spectrum, yielding 2500 spectra in total.
We extract quantities like $\lambda~L_{3000}$ and $\sigma_{MgII}$ (see section \ref{sec:measuring line width} and \ref{sec:FM}) and for four different cases; before or after reconstructing the spectra by applying PCA and before or after applying the radiation pressure correction (RPC).

\begin{figure}
\centering
  \includegraphics[scale=0.9,angle=0]{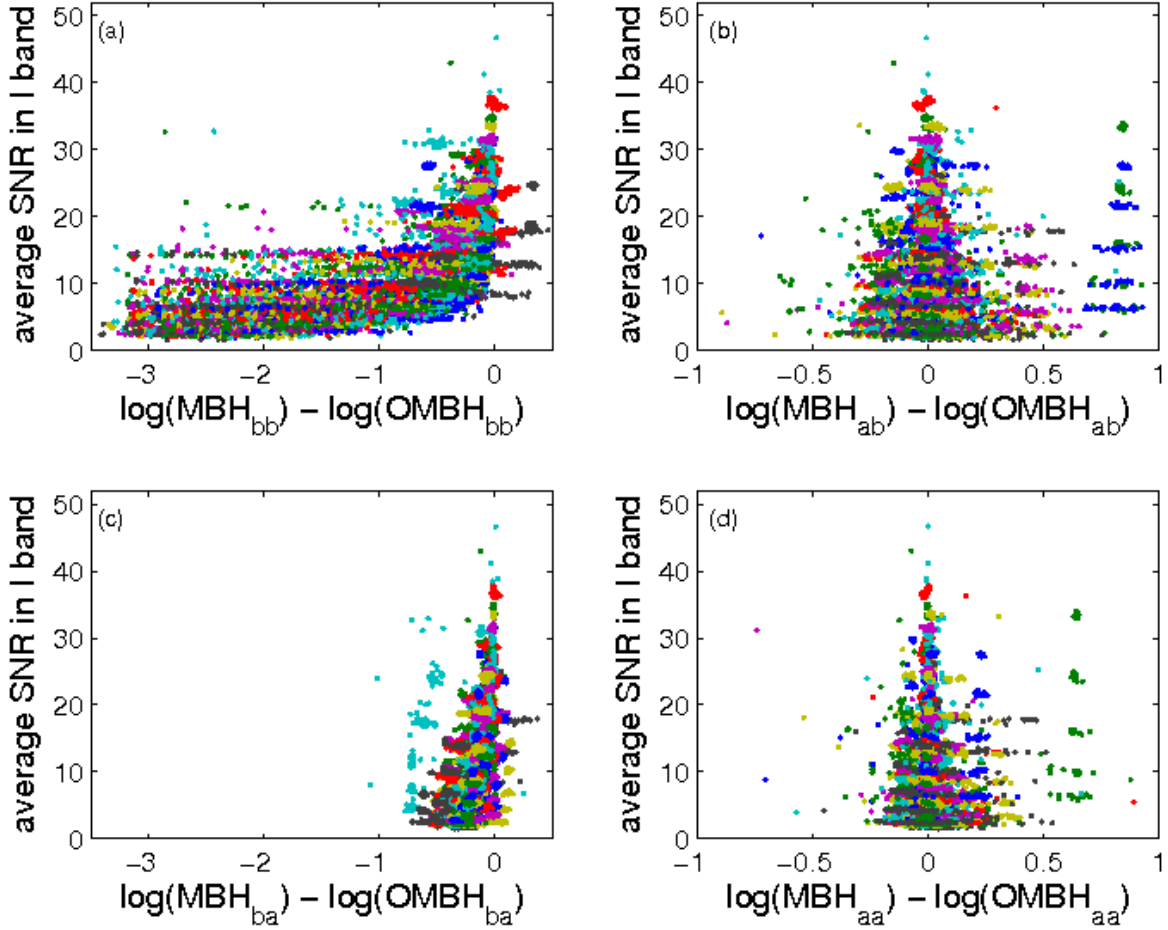}\\
    \caption{The log difference of the BH mass estimates for four scenarios. Different objects are represented in different colors. In all Figures, subscript \emph{a} stands for \emph{after} and subscript \emph{b} stands for \emph{before}. The first subscript states the PCA status while the second subscript states the Radiation Pressure Correction (RPC) status. Masses are estimated (a) before applying PCA and before RPC; (b) after applying PCA and before RPC; (c) before applying PCA but after RPC; (d) after applying PCA and after RPC. The cutoff on lower mass estimates in panel (c) is due to the asymptotic limit when $\sigma_{line}\rightarrow 0$. The presence of the light blue points at x-axis $\sim-0.7$ in this panel are due to the presence of the broad absorption line (BAL) in the \MgII\ window of an object. See the online version for color Figures.
}\label{fig:4in1_log_diff_MBH}
\end{figure}

Figure \ref{fig:4in1_log_diff_MBH} shows, as a function of the average SNR in the $I$ band, the differences in the logarithms of the black hole mass estimated after adding noise to spectra ($MBH$) and the mass estimated using original spectra with high SNR ($OMBH$) for four scenarios. As we can see from these Figures, the lower the SNR, the more the BH mass is uncertain with respect to the high SNR mass estimate. Comparing Figure \ref{fig:4in1_log_diff_MBH}a with Figure \ref{fig:4in1_log_diff_MBH}b shows an improvement in removing the systematic underestimation of BH masses when the spectra are reconstructed using PCA. Comparing Figure \ref{fig:4in1_log_diff_MBH}a with Figure \ref{fig:4in1_log_diff_MBH}c shows a cutoff on the lower side of the mass estimates in Figure \ref{fig:4in1_log_diff_MBH}c.
This lower limit is due to the asymptote approached when the estimated line dispersion is very small ($\lim_{\sigma \rightarrow 0} \textrm{\mbh}/\textrm{\msol}=\bar{g}\ell$) in the scenario when the RPC is applied. There were also three objects out of 100 selected with high SNR for which the results are catastrophic. One object is not being reconstructed properly due to an error on the % its redshift is around 0.9061, the header redshift is 0.9148, re-estimated by R\&H as 0.99358
redshift estimate, and two objects have a broad absorption line right in the \MgII\ window. Setting aside the 3\% catastrophic errors, the scatter is normally distributed around zero when PCA reconstruction is applied (see, e.g., Figure \ref{fig:4in1_log_diff_MBH}b and \ref{fig:4in1_log_diff_MBH}d).

In Figure \ref{fig:4in1_log_diff_MBH}a, if there was no SNR dependency then we would expect a Gaussian distribution around zero. The negative trend in mass difference in low SNR level shows the high sensitivity of the method to the SNR.
In Figure \ref{fig:4in1_log_diff_MBH}c, the non-symmetrical trend in low SNR emphasizes the SNR dependency too. This implies that using second moment of the line profile to estimate the $\sigma_{line}$ without PCA reconstruction is unbiased only for spectra with the very high SNR ($SNR > 30$) while after PCA reconstruction there is no such systematic bias with the SNR, making the PCA reconstruction a necessary step in precise estimation of BH mass algorithm.

\begin{figure}
\centering
  \includegraphics[scale=0.9,angle=0]{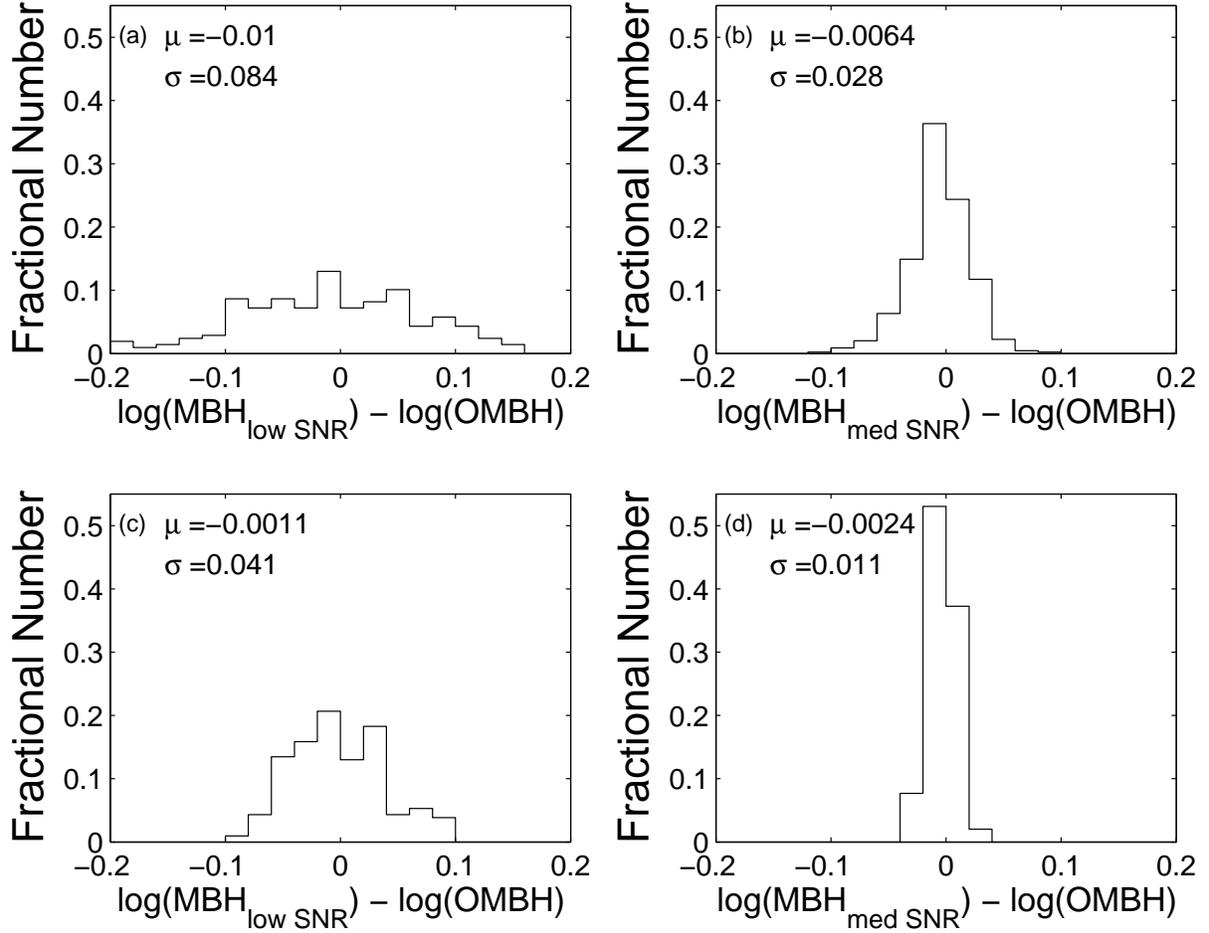}\\
  \caption{The distribution of the log difference of the BH mass estimates for two SNR bins: (a) and (c) $1.5 < SNR_{I} < 2.4$; (b) and (d) $18 < SNR_{I} < 23$. All plots are after applying PCA and for two scenarios: (a) and (b) before RPC; (c) and (d) after RPC. The mean ($\mu$) and the standard deviation ($\sigma$) of each distribution are reported in each panel. Comparing Figure (a) to (b) or comparing Figure (c) to (d) shows a factor of 3-3.7 improvement in mass estimates when high SNR is used. Comparing Figure (a) to (c) or comparing Figure (b) to (d) we have a factor of 2-2.5 improvement due using RPC.
  }\label{fig:4histogram_ab_RPC}
\end{figure}

Comparing the mass estimates from these different cases we can conclude that:
using PCA improves the mass estimation scatter to better than 0.4 dex for SNR as low as 1.5 or to better than 0.2 dex for SNR of 18 (see Figures \ref{fig:4histogram_ab_RPC}a and \ref{fig:4histogram_ab_RPC}b). This improvement is due to the sensitivity of the \FeII\ modeling process and the second moment measurement to both the SNR and the accuracy of the power-law estimates for quasar spectra.
Using PCA along with considering the effect of the radiation pressure improves the mass estimation scatter to better than 0.2 dex for SNR of around 1.5 (see Figures \ref{fig:4histogram_ab_RPC}c and \ref{fig:4histogram_ab_RPC}d).  Overall, BH mass estimates are around 50\% less scattered for either high or low SNR if we consider the effect of the radiation pressure (compare Figure \ref{fig:4in1_log_diff_MBH}b to \ref{fig:4in1_log_diff_MBH}d). However, BH masses are around 3 to 4 times less scattered in the high SNR regime than in the low SNR regime, either before (Figure \ref{fig:4in1_log_diff_MBH}b) or after the radiation correction (Figure \ref{fig:4in1_log_diff_MBH}d), but only if we reconstruct the spectra.

Denney et al.~(2009) have studied the effect of the SNR on reverberation BH mass estimates using both FWHM and line dispersion of quasar NGC~5548. They have reported a similar dependency on SNR as we see before reconstructing the spectra.

\section{Catalogue and Results}\label{sec:Catalogue results}

Here we present a catalogue including black hole mass estimates for three scenarios: black hole masses before applying PCA and before RPC ($MBH_{bb}$; see, e.g., Figure \ref{fig:3in1_MBH_z}a), BH masses after applying PCA but before RPC ($MBH_{ab}$; see, e.g., Figures  \ref{fig:3in1_MBH_z}b), and BH masses after applying PCA and after RPC ($MBH_{aa}$; see, e.g., Figures  \ref{fig:3in1_MBH_z}c). We present several quantities in this catalogue including the $\sigma_{MgII}$ and $\lambda~L_{3000}$ (see Table \ref{tab:table_catalogue_header} for more information about the catalogue format). In Figures  \ref{fig:3in1_MBH_z} and \ref{fig:3in1_cMBH_sigma} we present plots of different quantities corresponding to these three different scenarios. Comparing panel (b) against panel (a) of these Figures  represent the advantages of using PCA reconstruction while comparing panel (c) against panel (b) reveals the RPC.

The artificially low black hole mass estimates (especially in the high redshift regime) in Figure \ref{fig:3in1_MBH_z}a are mostly removed after PCA reconstruction in Figure \ref{fig:3in1_MBH_z}b and \ref{fig:3in1_MBH_z}c.
The masses were underestimated due to the decreased velocity range over which $\sigma_{line}$ could be reliably measured in low SNR spectra.  Forcing $\sigma_{line}$ to be measured over a fixed velocity range regardless of SNR might reduce the mass underestimation in low-SNR unreconstructed spectra, but only at the cost of much larger scatter on $\sigma_{line}$.
Figure \ref{fig:3in1_MBH_z}b shows that very massive black holes ($\gtrsim 10^{9.3} M_{\odot}$; the red points in the color version) were active as quasars at $z\gtrsim 1.5$.
However, a more reasonable true upper mass limit in our redshift range,
considering the errors which broaden the mass distribution,
is perhaps $10^{9.5} M_{\odot}$.

For first scientific uses of this catalogue,
see the BH spin evolution study of Rafiee \& Hall (2009)
and the sub-Eddington boundary study of Rafiee \& Hall (2011).

\subsection{Comparing with Shen et al. 2008}\label{sec:comparing_shen}
Shen et al.~(2008) have estimated black hole masses using the FWHMs of the \MgII, H$\beta$ and \CIV\ emission lines for objects in the SDSS DR5 quasar sample. We have extracted the FWHM \MgII\ BH masses, $L_{bol}$, and the FWHM of the same objects in our DR3 catalogue from Shen at al.~(2008).

Figure \ref{fig:2in1_L_LShen} (a and b) shows the relationship between our bolometric
luminosity and that of Shen et al. (2008) for the same objects.
There is a 0.1 dex scatter on the distribution of the log of the ratio
of $\lambda L_{3000}$ measurements (panel b) which is due to statistical errors.

The Mg\,{\sc ii} emission line width comparison shows a significant
difference between our estimates and those of Shen et al. (2008) (see Figure
\ref{fig:2in1_sig_FWHM_shen}, panels a to c).  The larger scatter in Figure
\ref{fig:2in1_sig_FWHM_shen}a is due to a
systematic underestimate of $\sigma$ in unreconstructed low-SNR spectra.  Figure
\ref{fig:2in1_sig_FWHM_shen}d shows the distribution of the ratio of our measured
line dispersion to the FWHM measured by Shen et al. (2008) in the same objects, relative
to the value of 1/2.3548 expected for a Gaussian.  At fixed FWHM, most quasar
Mg\,{\sc ii} emission lines have less flux in their wings than a Gaussian would.

Figure \ref{fig:3in1_M_Mshen} compares our $\log M_{BH}$ estimates with those of Shen et
al. (2008).  Figure \ref{fig:3in1_M_Mshen}a illustrates the effect of the systematic
underestimation of $M_{BH}$ in unreconstructed low-SNR spectra (Figure
\ref{fig:4in1_log_diff_MBH}a).  However,
even in PCA-reconstructed spectra (Figure \ref{fig:3in1_M_Mshen}c)
our BH mass estimates differ significantly from those of Shen et al. (2008).
We note from Figure \ref{fig:4in1_log_diff_MBH}b that there are no systematic errors
in our mass measurements from reconstructed spectra as a function of SNR;
we therefore have confidence in our measurements.
Figure \ref{fig:3in1_M_Mshen}b shows that the Shen et al. (2008) masses are overall
somewhat larger than ours.  Figure \ref{fig:3in1_M_Mshen}c illustrates that
relative to our masses from PCA-reconstructed spectra, the Shen et al. (2008) masses are,
on average, overestimated at higher masses and underestimated at lower masses.

The \MgII\ emission line width comparison shows a significant difference between our estimates and those of Shen et al.~(2008) (see Figures \ref{fig:2in1_sig_FWHM_shen}a to \ref{fig:2in1_sig_FWHM_shen}c). However, we note from Figure \ref{fig:4in1_log_diff_MBH}b that there is no systematic errors in our mass measurements as a function of SNR.

\subsection{Malmquist Bias and the true distribution of BH masses}\label{sec:Malmquist Bias}
A Malmquist bias ($\delta_M$; Eddington~1913; Malmquist~1922), i.e., the difference between the mean values of the distributions of the true BH masses and the observed BH masses, is estimated for each of our three scenarios. This Malmquist bias, which is a selection effect, exists since we are using a flux-limited sample of quasars and since the mass function of quasars is very steep.
There are more quasars with small masses being scattered to large masses in our sample by random uncertainties than there are quasars with large masses being scattered to small masses in our sample, leading to overestimated BH masses on average.

We use the generalized idea of the bias at fixed observed luminosity described in detail by Shen et al. (2008). This bias is estimated as $\delta_M=-\gamma_{M} \sigma_{\Upsilon}^{2}\textrm{ln}(10)/ (1+C_{2})^{2}$ where $C_{2}$ is the slope of the assumed relationship between $\log \textrm{\mbh}_{,true}$ and $\langle\log\Upsilon\rangle$, where $\Upsilon=L_{bol}/L_{Edd}$ is the Eddington ratio and $\gamma_{M}$ is the slope when we assume a power-law distribution for the underlying
true BH masses as $N(\textrm{\mbh})\propto\textrm{\mbh}^{\gamma_{M}}$. The $\sigma_{\Upsilon}$ is defined as the dispersion of the true distribution of the Eddington ratio (see Shen et al.~2008 for more details). Shen et al. has modeled the underlying distribution of the true BH masses and estimated $C_{2}=0.3$ and $\gamma_{M}=-2.6$ for the redshift bin of $0.7 < z < 1.0$. We assume the same distribution of true BH masses and estimate the $\sigma_{\Upsilon}$ for all three of our scenarios. We then estimate the Malmquist bias ($\delta_M$) of the three scenarios to be $\delta_M(bb)\simeq0.44$, $\delta_M(ab)\simeq0.40$, and $\delta_M(aa)\simeq0.11$. However, we do not apply these estimated Malmquist biases on reported BH masses in the catalogue,
as the calculation of the Malmquist bias is model dependent and users of the
catalogue may wish to apply a differently calculated correction.
The Malmquist bias for the case after applying both PCA and RPC is very small since the BH mass distribution is dominated by the narrow distribution of the luminosity due to the radiation pressure correction (see Equation \ref{equ:RPC}).

%=======================================================================================================
\section{Conclusions}\label{sec:conclusion}
\begin{figure}
\centering
  \includegraphics[scale=0.9,angle=0]{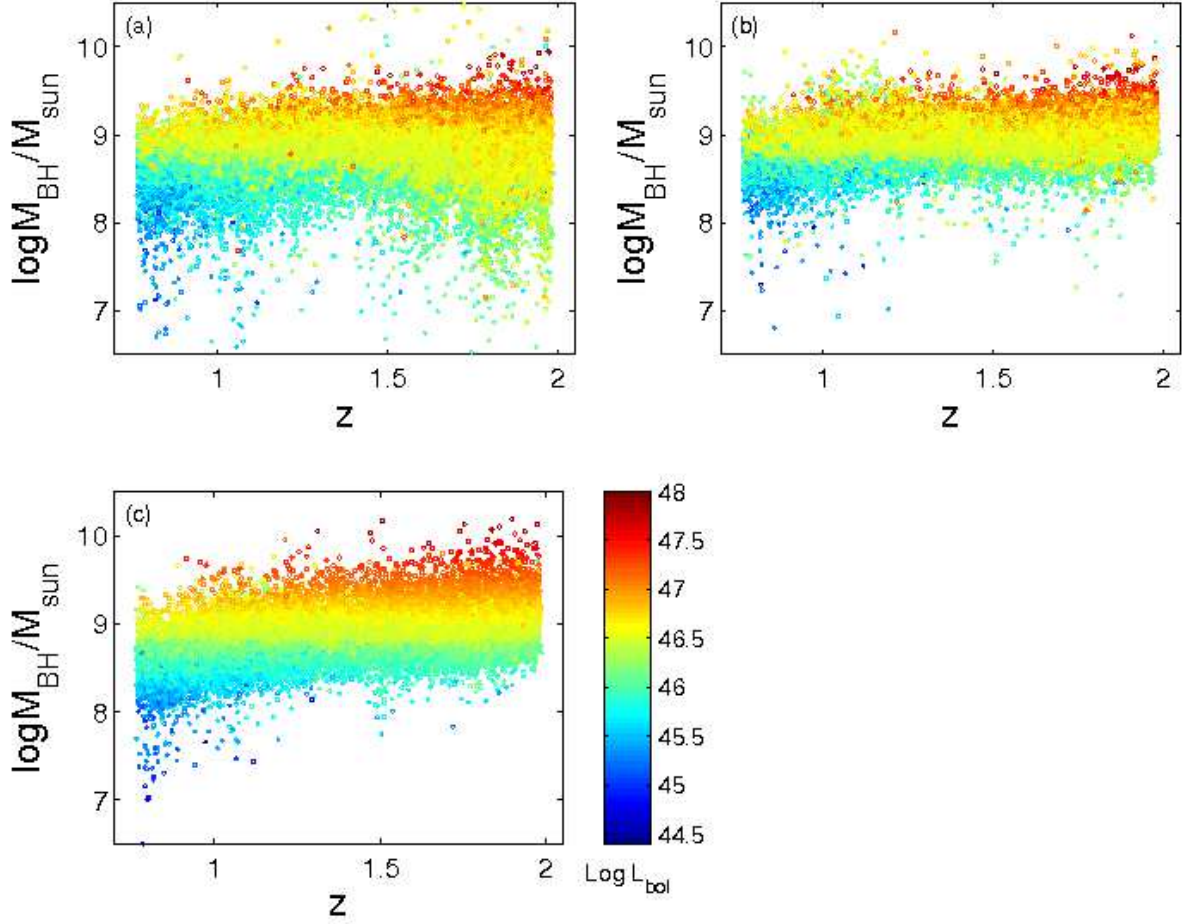}\\
  \caption{The $\log~\textrm{\mbh}$ versus redshift plane for three scenarios. (a) Before applying PCA and RPC. (b) After applying PCA and before RPC. (c) After applying PCA and RPC. The lack of low mass BHs at high $z$ and high mass BHs at low $z$ is due to detection and saturation limits. BH masses between $10^{8.5}$ and $10^{9.3}$ are distributed all over our redshift range. See the online version for color Figures.
  }\label{fig:3in1_MBH_z}
\end{figure}

\begin{figure}
\centering
  \includegraphics[scale=0.9,angle=0]{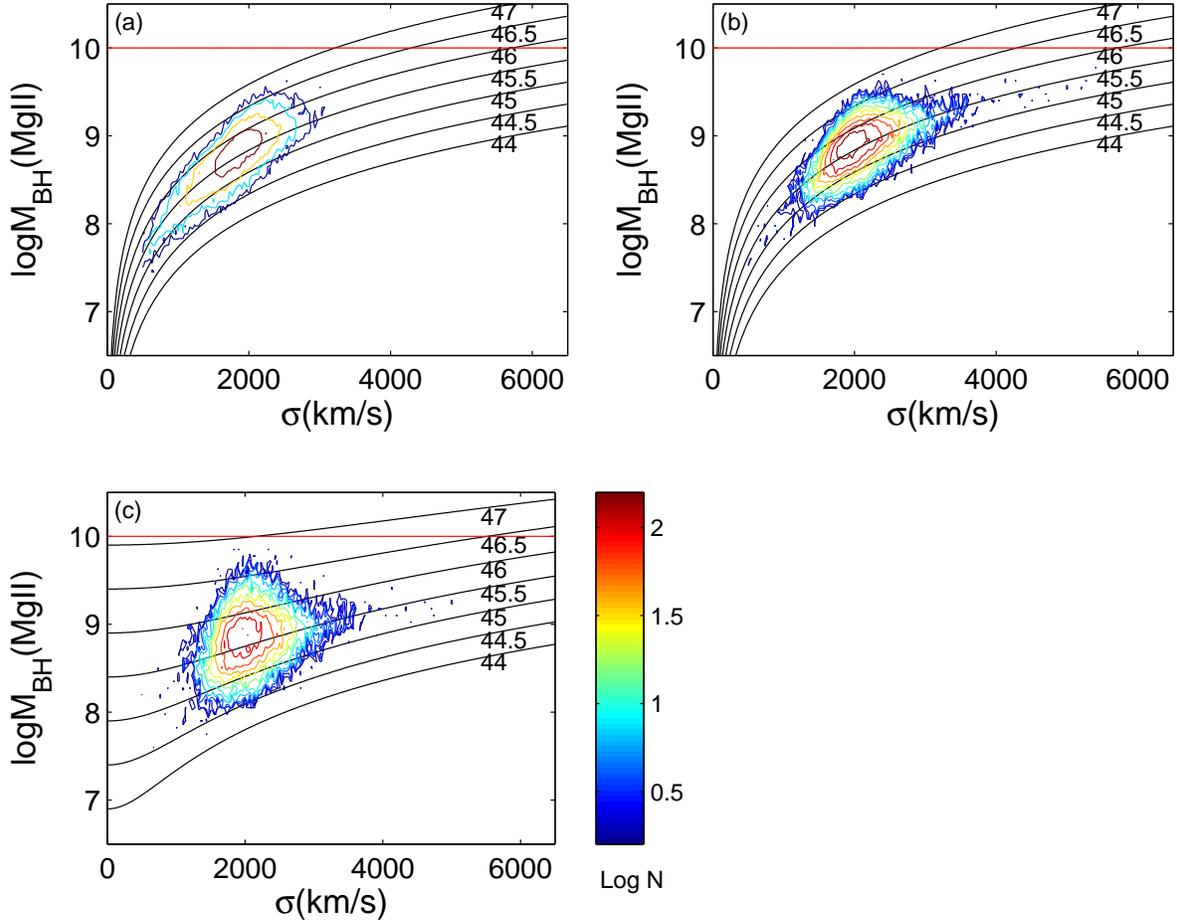}\\
  \caption{Contours of $\log~\textrm{\mbh}$ in solar units versus $\sigma_{MgII}$ plane for three scenarios. (a) Before applying PCA and RPC. (b) After applying PCA and before RPC. (c) After applying PCA and RPC. The black curves show the $\log~\textrm{\mbh}$ versus $\sigma_{MgII}$ for a fixed $\log~L_{bol}$ ranging from 44 to 47. The red line $\log~\textrm{\mbh}=10$ represents the upper mass limit suggested by Shen et al. (2008). See the online version for color Figures.}\label{fig:3in1_cMBH_sigma}
\end{figure}

\begin{figure}
\centering
  \includegraphics[scale=0.9,angle=0]{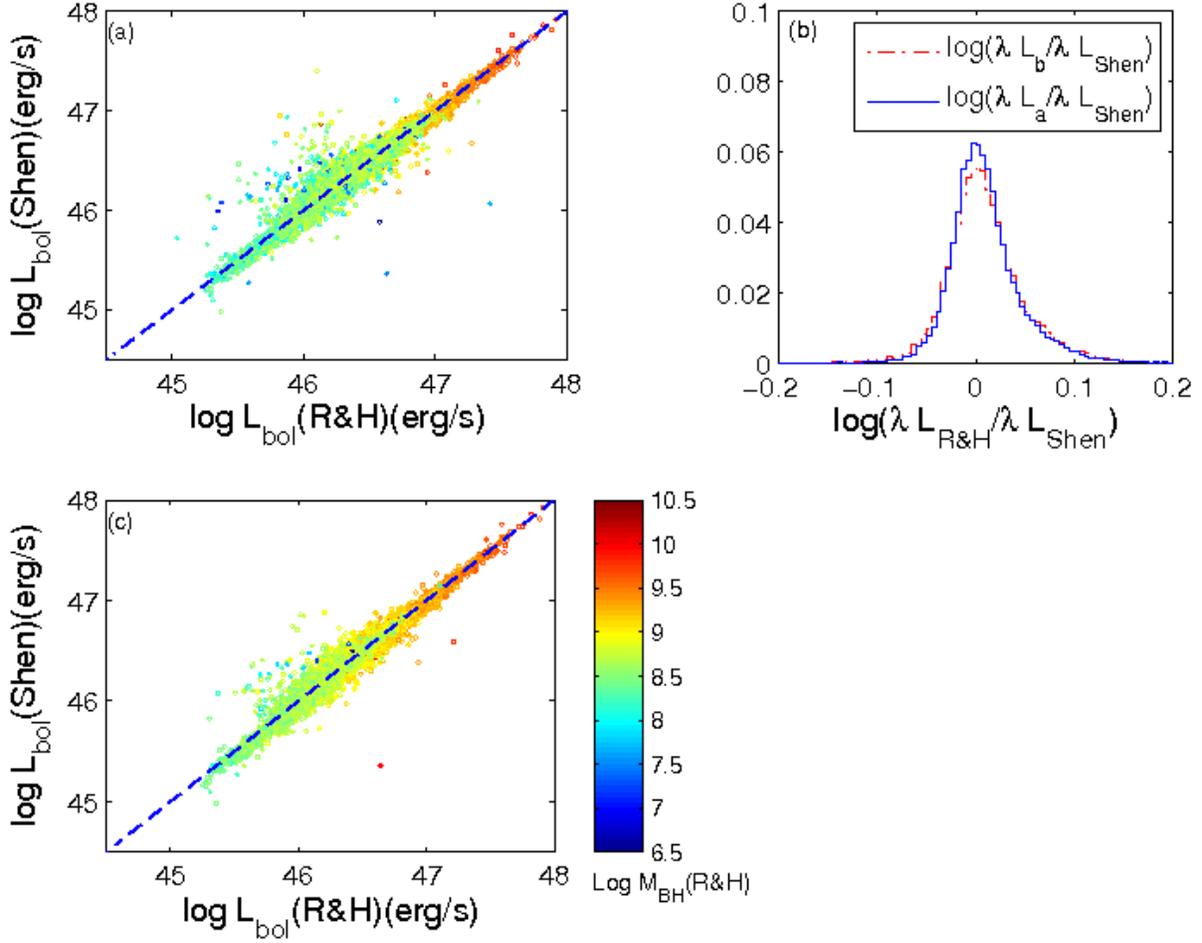}\\
  \caption{${\rm Log} L_{bol}(Shen)$ versus $\log L_{bol}(R\&H)$ for two scenarios. (a) Before applying PCA and RPC. (c) After applying PCA and before RPC. $L_{b}$ and $L_{a}$ are Luminosities before and after applying PCA (but before applying RPC) respectively. Both panels show consistency between Shen et al. (2008) and this study; in both, there is a 0.1 dex scatter on the distribution (panel b) due to statistical errors.  See the online version for color Figures.}\label{fig:2in1_L_LShen}
\end{figure}

\begin{figure}
\centering
  \includegraphics[scale=0.9,angle=0]{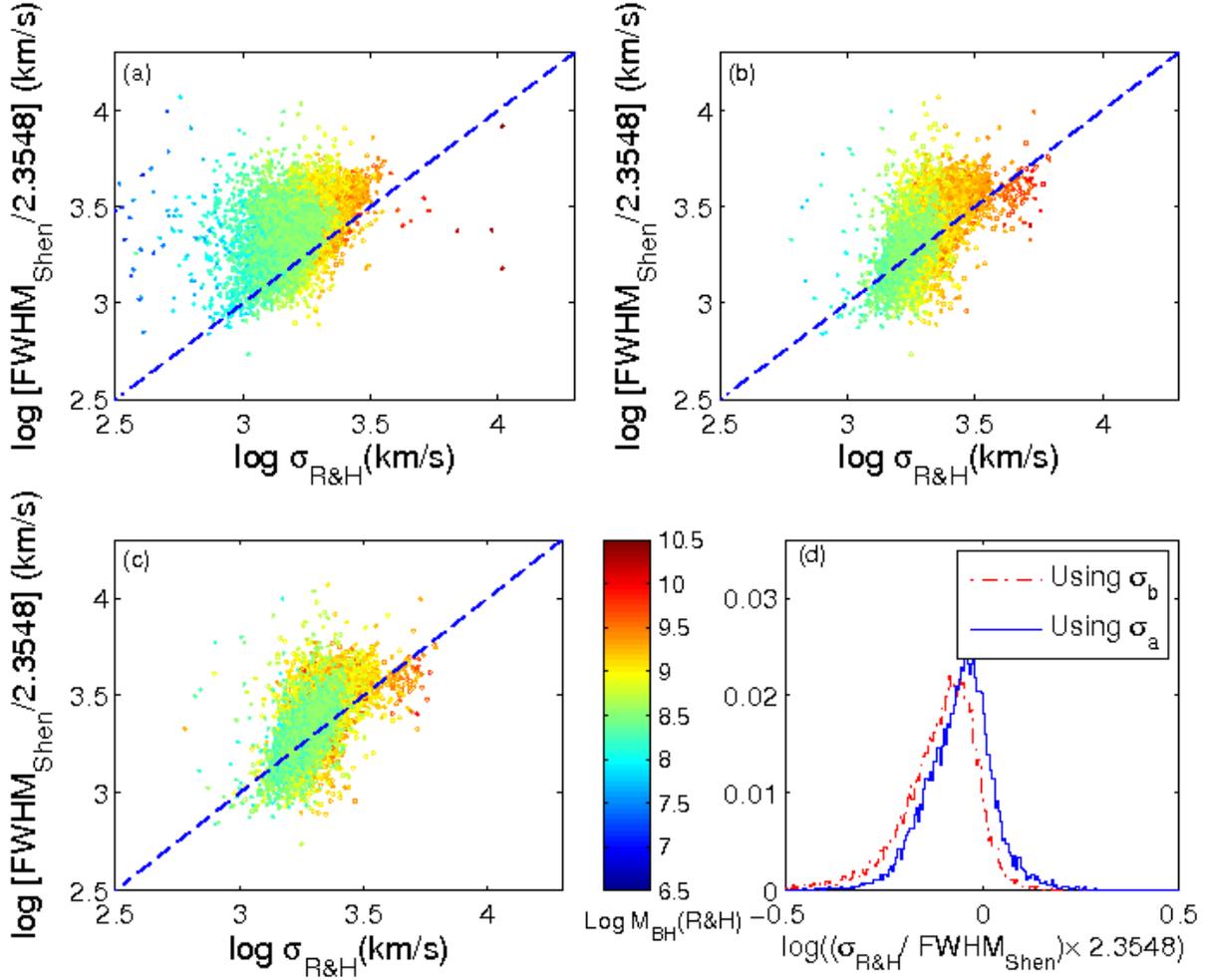}\\
 \caption{${\rm Log} (\textrm{FWHM}_{Shen}/2.3548)$ versus $\log \sigma_{R\&H}$ for three scenarios. (a) Before applying PCA and RPC. (b) After applying PCA and before RPC. (c) After applying PCA and RPC. Although $\sigma_a$ (the line dispersion after applying PCA) is the same in both (b) and (c) they generate different \mbh\ distributions since the scaling relationship is different. $\sigma_b$ is the line dispersion before applying PCA.  Panel (d) shows the distribution of the ratio of our measured line dispersion to the FWHM measured by Shen et al. (2008) in the same objects, relative to the value of 1/2.3548 expected for a Gaussian.  See the online version for color Figures.}\label{fig:2in1_sig_FWHM_shen}
\end{figure}

\begin{figure}
\centering
  \includegraphics[scale=0.9,angle=0]{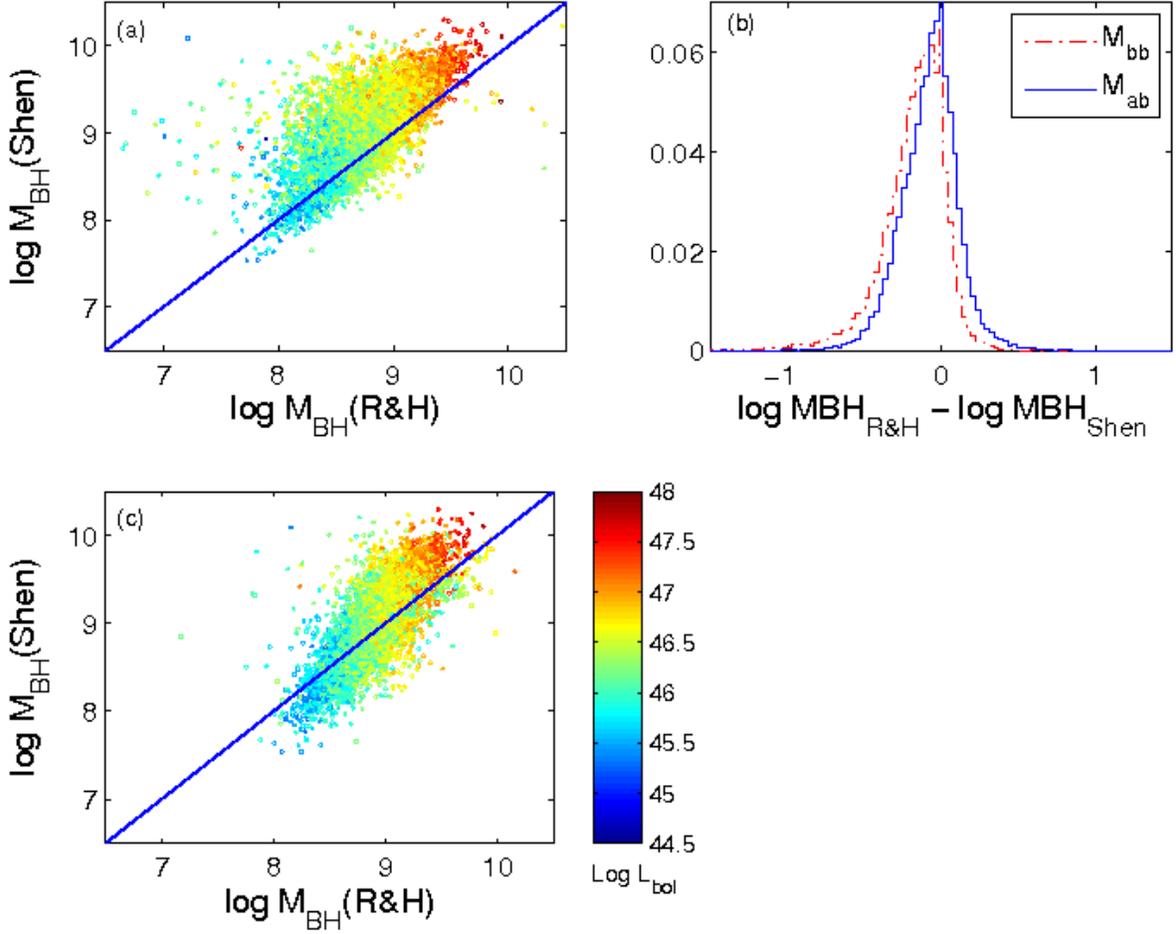}\\
  \caption{${\rm Log} \textrm{\mbh}(Shen)$ versus $\log \textrm{\mbh}(R\&H)$ in solar units for two scenarios. (a) Before applying PCA and RPC. (c) After applying PCA and before RPC.  Panel (b) shows that the Shen et al. (2008) masses are overall somewhat larger than ours; $M_{bb}$ and $M_{ab}$ are our \mbh\ values computed before and after applying PCA (but before applying RPC) respectively. The steeper slope in panel (c) is due to the same effect on $\sigma_{MgII}$ measurement as in Figure \ref{fig:2in1_sig_FWHM_shen}, meaning that relative to our masses, the high mass BHs are overestimated in Shen et al. (2008) while the low mass BHs are underestimated. See the online version for color Figures.}\label{fig:3in1_M_Mshen}
\end{figure}

We have measured virial BH masses and radiation pressure adjusted virial BH
masses for 27,602 quasars at $0.7<z<2.0$ in the SDSS DR3 quasar catalogue,
and have made our mass estimates available to the community.
We have used the second moment of the \MgII\ emission line profile
to estimate the line width and have calibrated a virial mass estimator
using it and the monochromatic luminosity at 3000{\AA}.  The virial BH masses
are typically in the range $10^8-10^{9.5} \textrm{\msol}$ for our quasar
sample (Figures \ref{fig:3in1_MBH_z} and \ref{fig:3in1_cMBH_sigma}).

We have reconstructed the quasar spectra using Principal Component Analysis to increase the number of quasars for which reliable masses can be measured and to eliminate systematic biases due to the presence of noise. We have tested the reliability and biases of the mass measurements as a function of SNR before and after this PCA reconstruction.
Our noise simulation shows no systematic bias in measured BH masses as a function of SNR (section \ref{sec:Noise simulation}). Using PCA reconstructed spectra for measuring BH masses reduces the intrinsic scatter in the BH mass estimates.

For a subsample of Shen et al.~(2008) quasars in common with our sample, we have compared the estimated bolometric luminosity, the line width, and the BH mass. The bolometric luminosities are mutually consistent, with 0.1 dex statistical scatter. However, with respect to this study, Shen et al.~(2008) have overestimated the line width in the case of the broadest emission lines and underestimated the line width in the case of the narrowest emission lines. This difference in line width measurement has affected the BH mass estimates such that those of the most massive BHs in Shen et al. are overestimated and those of the least massive BHs are underestimated with respect to this study.
We believe our black hole mass estimates are accurate.
Improved understanding of black hole masses is necessary for understanding properties of the distribution of quasars in parameter space, such as the sub-Eddington boundary (Steinhardt \& Elvis 2010; Rafiee \& Hall 2011).

%=======================================================================================================

\acknowledgments
We thank C.-W. Yip, D. Vanden Berk, and T. Boroson for sharing methods and expertise.  PBH and AR are supported in part by NSERC.
Funding for the SDSS and SDSS-II was provided by the Alfred P. Sloan Foundation, the Participating Institutions, the National Science Foundation, the U.S. Department of Energy, the National Aeronautics and Space Administration, the Japanese Monbukagakusho, the Max Planck Society, and the Higher Education Funding Council for England. The SDSS was managed by the Astrophysical Research Consortium for the Participating Institutions.

%=======================================================================================================

%\pagebreak
%\clearpage
%=======================================================================================================

%-----------------------------------------

%--------------------------
%--------------------------
%--------------------------
%--------------------------
%--------------------------
%---------------------------
%---------------------------
%-------------------------
%---------------------------

%=======================================================================================================
%\newpage
%\begin{landscape}
\begin{deluxetable}{lcccccccccc}
\rotate
%\tabletypesize{\scriptsize}
\tabletypesize{\tiny}
\tablecaption{Calibration Data. \label{tab:tbl-1}} \tablewidth{0pt} \tablehead{
 \colhead{Object}
&\colhead{DataID}
&\colhead{$\sigma_{MgII}$}
&\colhead{$\lambda L_{3000}/10^{44}$}
&\colhead{$<\tau>_{H\alpha,\beta,\gamma}$}
&\colhead{$M_{BH}$}
&\colhead{flag 1}
&\colhead{Virial $M_{BH}$}
&\colhead{flag 2}
&\colhead{Adjusted Virial $M_{BH}$}
&\colhead{$\sigma^{*}_{bulge}$}\\
 \colhead{ }
&\colhead{ }
&\colhead{km/s}
&\colhead{erg/s}
&\colhead{days}
&\colhead{$M_{\odot}\times10^{6}$}
&\colhead{ }
&\colhead{$M_{\odot}\times10^{6}$}
&\colhead{ }
&\colhead{$M_{\odot}\times10^{6}$}
&\colhead{km/s}\\
\colhead{1}&\colhead{2}&\colhead{3}&\colhead{4}&\colhead{5}&\colhead{6}&\colhead{7}&\colhead{8}&\colhead{9}&\colhead{10}&\colhead{11}
}\startdata
3C120		&	LWP04500	&		&		&								&								&			&					&		&				 &					\\
		&	LWR16609	&		&		&								&								&			&					&		&				 &					\\
		&		&	3384.0	&	1.7472	&	38.10	$^{	+	21.30	}_{	-15.30	}$	&	55.5	$^{	+	31.4	}_{	-22.5	}$	&	-1		&	87.0	 $\pm$	3.7		&	1	&	222	$\pm$	10	&	162.0	$\pm$	24.0		\\
3C390.3		&	LWP16776	&		&		&								&								&			&					&		&				 &					\\
		&	LWP03808	&		&		&								&								&			&					&		&				 &					\\
		&	LWP17245	&		&		&								&								&			&					&		&				 &					\\
		&	LWP18382	&		&		&								&								&			&					&		&				 &					\\
		&	LWP17259	&		&		&								&								&			&					&		&				 &					\\
		&	LWP17478	&		&		&								&								&			&					&		&				 &					\\
		&		&	19533.8	&	0.5397	&	23.60	$^{	+	4.38	}_{	-4.74	}$	&	287.0	$^{	+	64.0	}_{	-64.0	}$	&	-1	 \tablenotemark{d}	&	1610	$\pm$	220		&	-1	&	3870	$\pm$	540	&	240.0	$\pm$	36.0		\\
Akn120(Ark120)	\tablenotemark{\dagger}	&	Y29E0106T	&	1377.7	&	0.0417	&	39.05	$^{	+	4.16	}_{	-4.95	}$	&	150.0	$^{	+	19.0	 }_{	-19.0	}$	&	1		&	2.2	$\pm$	4.0		&	1	&	5.7	$\pm$	11	&	239.0	$\pm$	36.0		\\
Fairall9		&	LWP24094	&		&		&								&								&			&					&		 &				&					\\
		&	LWP24095	&		&		&								&								&			&					&		&				 &					\\
		&	LWP06492	&		&		&								&								&			&					&		&				 &					\\
		&	LWP19270	&		&		&								&								&			&					&		&				 &					\\
		&	LWP19554	&		&		&								&								&			&					&		&				 &					\\
		&	LWP21475	&		&		&								&								&			&					&		&				 &					\\
		&	LWP21474	&		&		&								&								&			&					&		&				 &					\\
		&	LWP21473	&		&		&								&								&			&					&		&				 &					\\
		&	LWP19915	&		&		&								&								&			&					&		&				 &					\\
		&		&	1797.7	&	3.4198	&	17.40	$^{	+	3.20	}_{	-4.30	}$	&	255.0	$^{	+	56.0	}_{	-56.0	}$	&	1		&	34.4	 $\pm$	0.8		&	-1	&	109.6	$\pm$	3.0	&	\nodata				\\\\
Mrk110		&	LWP12760	&		&		&								&								&			&					&		&				 &					\\
		&	LWP12761	&		&		&								&								&			&					&		&				 &					\\
		&		&	1431.5	&	0.5864	&	24.37	$^{	+	4.63	}_{	-4.49	}$	&	25.1	$^{	+	6.1	}_{	-6.1	}$	&	1		&	9.0	$\pm$	 1.2		&	1	&	26.3	$\pm$	4.0	&	86.0	$\pm$	13.0		\\\\
Mrk335		&	LWR11470	&		&		&								&								&			&					&		&				 &					\\
		&	LWR03512	&		&		&								&								&			&					&		&				 &					\\
		&	LWR09917	&		&		&								&								&			&					&		&				 &					\\
		&	LWR11947	&		&		&								&								&			&					&		&				 &					\\
		&	LWR14554	&		&		&								&								&			&					&		&				 &					\\
		&	LWR14555	&		&		&								&								&			&					&		&				 &					\\
		&	LWP19466	&		&		&								&								&			&					&		&				 &					\\
		&	LWP19123	&		&		&								&								&			&					&		&				 &					\\
		&	LWP16819	&		&		&								&								&			&					&		&				 &					\\
		&		&	2028.7	&	1.3369	&	15.27	$^{	+	3.88	}_{	-3.34	}$	&	14.2	$^{	+	3.7	}_{	-3.7	}$	&	1		&	27.4	 $\pm$	1.5		&	-1	&	76.3	$\pm$	4.9	&	\nodata				\\\\
Mrk509		&	LWR13716	&		&		&								&								&			&					&		&				 &					\\
		&	LWR01636	&		&		&								&								&			&					&		&				 &					\\
		&	LWP14218	&		&		&								&								&			&					&		&				 &					\\
		&	LWR01309	&		&		&								&								&			&					&		&				 &					\\
		&	LWP14534	&		&		&								&								&			&					&		&				 &					\\
		&	LWP10829	&		&		&								&								&			&					&		&				 &					\\
		&	LWP15474	&		&		&								&								&			&					&		&				 &					\\
		&	LWR06219	&		&		&								&								&			&					&		&				 &					\\
		&	LWP18844	&		&		&								&								&			&					&		&				 &					\\
		&	LWP18786	&		&		&								&								&			&					&		&				 &					\\
		&	LWP18785	&		&		&								&								&			&					&		&				 &					\\
		&		&	1977.1	&	4.1095	&	79.60	$^{	+	6.10	}_{	-5.40	}$	&	143.0	$^{	+	12.0	}_{	-12.0	}$	&	1		&	45.6	 $\pm$	0.8		&	-1	&	142.0	$\pm$	3.2	&	\nodata				\\\\
Mrk590		&	LWP19577	&		&		&								&								&			&					&		&				 &					\\
		&	LWR13721	&		&		&								&								&			&					&		&				 &					\\
		&		&	2246.2	&	0.7761	&	24.23	$^{	+	2.16	}_{	-2.01	}$	&	47.5	$^{	+	7.4	}_{	-7.4	}$	&	1		&	25.6	 $\pm$	2.5		&	1	&	67.5	$\pm$	7.1	&	194.0	$\pm$	20.0		\\\\
Mrk79		&	LWR06141	&		&		&								&								&			&					&		&				 &					\\
		&	LWR01320	&		&		&								&								&			&					&		&				 &					\\
		&		&	1621.4	&	0.4975	&	14.38	$^{	+	4.02	}_{	-3.80	}$	&	52.4	$^{	+	14.4	}_{	-14.4	}$	&	1		&	10.7	 $\pm$	1.6		&	1	&	29.5	$\pm$	5.0	&	130.0	$\pm$	20.0		\\\\
Mrk817		&	LWR11936	&	2054.0	&	1.0361	&	19.05	$^{	+	2.48	}_{	-2.41	}$	&	49.4	$^{	+	7.7	}_{	-7.7	}$	&	1		&	 24.7	$\pm$	1.8		&	1	&	67.5	$\pm$	5.5	&	142.0	$\pm$	21.0		\\
NGC3783		&	LWP23003	&		&		&								&								&			&					&		&				 &					\\
		&	LWP23318	&		&		&								&								&			&					&		&				 &					\\
		&	LWP23311	&		&		&								&								&			&					&		&				 &					\\
		&	LWP23298	&		&		&								&								&			&					&		&				 &					\\
		&	LWP23289	&		&		&								&								&			&					&		&				 &					\\
		&	LWP23280	&		&		&								&								&			&					&		&				 &					\\
		&	LWP23094	&		&		&								&								&			&					&		&				 &					\\
		&	LWP23075	&		&		&								&								&			&					&		&				 &					\\
		&	LWP23073	&		&		&								&								&			&					&		&				 &					\\
		&	LWP23035	&		&		&								&								&			&					&		&				 &					\\
		&	LWP22902	&		&		&								&								&			&					&		&				 &					\\
		&	LWP22876	&		&		&								&								&			&					&		&				 &					\\
		&	LWP22845	&		&		&								&								&			&					&		&				 &					\\
		&	LWP22827	&		&		&								&								&			&					&		&				 &					\\
		&	LWP22826	&		&		&								&								&			&					&		&				 &					\\
		&	LWP23074	&		&		&								&								&			&					&		&				 &					\\
		&		&	1202.7	&	0.2696	&	10.20	$^{	+	3.30	}_{	-2.30	}$	&	29.8	$^{	+	5.4	}_{	-5.4	}$	&	1		&	4.3	$\pm$	 1.2		&	1	&	12.5	$\pm$	4.1	&	95.0	$\pm$	10.0		\\\\
NGC4051		&	LWP23153	&		&		&								&								&			&					&		&				 &					\\
		&	LWP19265	&		&		&								&								&			&					&		&				 &					\\
		&	LWP27298	&		&		&								&								&			&					&		&				 &					\\
		&	LWP27297	&		&		&								&								&			&					&		&				 &					\\
		&	LWP24347	&		&		&								&								&			&					&		&				 &					\\
		&	LWP11100	&		&		&								&								&			&					&		&				 &					\\
		&		&	1131.4	&	0.0080	&	5.80	$^{	+	2.60	}_{	-1.80	}$	&	1.9	$^{	+	0.78	}_{	-0.78	}$	&	-1		&	0.7	$\pm$	 6.2		&	1	&	1.6	$\pm$	16.0	&	84.0	$\pm$	9.0		\\
NGC5548		&	LWP24683	&		&		&								&								&			&					&		&				 &					\\
		&	LWP25088	&		&		&								&								&			&					&		&				 &					\\
		&	LWP25452	&		&		&								&								&			&					&		&				 &					\\
		&	LWP25440	&		&		&								&								&			&					&		&				 &					\\
		&	LWP25409	&		&		&								&								&			&					&		&				 &					\\
		&	LWP25464	&		&		&								&								&			&					&		&				 &					\\
		&	LWP25422	&		&		&								&								&			&					&		&				 &					\\
		&	LWP25483	&		&		&								&								&			&					&		&				 &					\\
		&	LWP25522	&		&		&								&								&			&					&		&				 &					\\
		&	LWP25514	&		&		&								&								&			&					&		&				 &					\\
		&	LWP25496	&		&		&								&								&			&					&		&				 &					\\
		&		&	1971.3	&	0.3769	&	16.81	$^{	+	0.97	}_{	-0.98	}$	&	67.1	$^{	+	2.6	}_{	-2.6	}$	&	1		&	13.7	 $\pm$	2.7		&	1	&	35.9	$\pm$	7.7	&	183.0	$\pm$	27.0		\\
NGC7469	\tablenotemark{\dagger}	&	Y3B4010CT	&	1798.8	&	0.5957	&	4.54	$^{	+	0.64	}_{	-0.68	}$	&	12.2	$^{	+	1.4	}_{	-1.4	 }$	&	1		&	14.4	$\pm$	1.8		&	1	&	39.2	$\pm$	5.5	&	152.0	$\pm$	16.0		\\
PG0844+349		&	LWP12206	&	1047.1	&	4.6751	&	31.99	$^{	+	13.45	}_{	-12.64	}$	&	92.4	$^{	+	38.1	}_{	-38.1	}$	&	-1		 &	13.6	$\pm$	0.2		&	-1	&	69.9	$\pm$	1.7	&	\nodata				\\
PG1211+143		&	LWP13628	&		&		&								&								&			&					&		 &				&					\\
		&	LWP07603	&		&		&								&								&			&					&		&				 &					\\
		&	LWP07223	&		&		&								&								&			&					&		&				 &					\\
		&	LWP23329	&		&		&								&								&			&					&		&				 &					\\
		&	LWP19386	&		&		&								&								&			&					&		&				 &					\\
		&		&	958.4	&	8.8465	&	103.94	$^{	+	17.39	}_{	-23.64	}$	&	146.0	$^{	+	44.0	}_{	-44.0	}$	&	1		&	15.7	 $\pm$	0.1		&	-1	&	108.0	$\pm$	1.5	&	\nodata				\\
PG1226+023	\tablenotemark{\dagger}	&	Y0G4020FT	&	1816.2	&	85.9274	&	352.99	$^{	+	32.10	}_{	-37.45	}$	&	886.0	$^{	+	187.0	 }_{	 -187.0	}$	&	1		&	175.8	$\pm$	0.2		&	-1	&	1104	$\pm$	1.6	&	\nodata				\\
PG1229+204		&	LWR13136	&		&		&								&								&			&					&		 &				&					\\
		&	LWR16071	&		&		&								&								&			&					&		&				 &					\\
		&		&	1558.2	&	3.1430	&	33.33	$^{	+	14.59	}_{	-12.35	}$	&	73.2	$^{	+	35.2	}_{	-35.2	}$	&	-1		&	24.8	 $\pm$	0.6		&	1	&	84.4	$\pm$	2.6	&	162.0	$\pm$	32.0	\tablenotemark{\ddag}	\\\\
PG1351+640	\tablenotemark{\dagger}	&	Y0P80305T	&		&		&								&								&			&					 &		&				&					\\
		&	O65616010	&		&		&								&								&			&					&		&				 &					\\
		&		&	1995.4	&	6.9307	&	\nodata							&	\nodata							&	-1		&	60.3	$\pm$	0.7		 &	-1	&	199.7	$\pm$	2.8	&	\nodata				\\\\
PG1411+442	\tablenotemark{\dagger}	&	Y11U0103T	&	1700.3	&	3.9038	&	101.58	$^{	+	31.00	}_{	-28.06	}$	&	443.0	$^{	+	146.0	 }_{	 -146.0	}$	&	1		&	32.8	$\pm$	0.6		&	-1	&	109.8	$\pm$	2.7	&	\nodata				\\
PG2130+099		&	LWR04610	&		&		&								&								&			&					&		 &				&					\\
		&	LWP03568	&		&		&								&								&			&					&		&				 &					\\
		&		&	1478.9	&	3.9699	&	177.22	$^{	+	19.94	}_{	-12.68	}$	&	457.0	$^{	+	55.0	}_{	-55.0	}$	&	1		&	25.1	 $\pm$	0.5		&	1	&	91.7	$\pm$	2.3	&	172.0	$\pm$	46.0	\tablenotemark{\ddag}	\\
\enddata
\tablecomments{Column 1 details the name of the objects and its resources. Column 2 are the IUE or HST spectra. Columns 3 and 4 are the \MgII\ line dispersion in units of km s$^{-1}$ and the monochromatic luminosity at 3000{\AA} respectively. Column 5 is the average time-lag in units of light-days from Peterson et al. 2004. Column 6 is the reverberation mapping BH mass estimates from Peterson et al. 2004. Columns 8 and 10 are the BH mass estimates for two scenarios, before or after applying the radiation correction, respectively. Columns 7 and 9 are the outlier flag in regression analysis; flag 1 for Virial $M_{BH}$ and flag 2 for Adjusted Virial $M_{BH}$. Column 11 is the stellar velocity dispersion of the bulge from Onken et al. 2004 and Dasyra et al. 2007.}
\tablenotetext{\dagger}{HST data.}
\tablenotetext{\ddag}{Bulge Velocity dispersion extracted from Dasyra et al. 2007.}
\tablenotetext{d}{Double peak line.}
\end{deluxetable}
%\end{landscape}
%\clearpage
%=======================================================================================================
%\newpage
%\tabletypesize{\scriptsize}
\begin{deluxetable}{lccc}
%% This is the title of the table.
\tablewidth{0pt}
\begin{minipage}{4.2in}
\tablecaption{$\log(\tau)=\alpha\log(\ell)+ \beta$, Fitting Results.\label{tab:fitting-results}}
\end{minipage}
\tablehead{\colhead{Fitting Method} & \colhead{Slope($\alpha$)} & \colhead{Intercept($\beta$)}& \colhead{Scatter}
}
\startdata
FITEXY	&	0.55	$\pm$	0.02	&	1.49	$\pm$	0.02	&	0	\\
FITEXY	&	0.47	$\pm$	0.12	&	1.46	$\pm$	0.09	&	0.35	\\
\hline											
BCES(Y$|$X)	&	0.48	$\pm$	0.16	&	1.46	$\pm$	0.10	&	0	\\
bootstrap	&	0.51	$\pm$	0.20	&	1.43	$\pm$	0.10	&	0	\\
\hline											
BCES Bisector	&	0.63	$\pm$	0.11	&	1.44	$\pm$	0.10	&	0	\\
bootstrap	&	0.66	$\pm$	0.18	&	1.43	$\pm$	0.11	&	0	\\
\hline												
MCMC	\tablenotemark{\dagger}	&	0.40	$\pm$	0.10	&	1.46	$\pm$	0.08	&	0.33	\\
\enddata
\tablenotetext{\dagger}{Markov Chain Monte-Carlo Simulation}
\end{deluxetable}
%\clearpage
%=======================================================================================================
% Table generated by Excel2LaTeX from sheet 'header3'
%\newpage
\begin{deluxetable}{lcccl}
\tabletypesize{\scriptsize}
%\rotate
\tablecaption{Catalogue Format. \label{tab:table_catalogue_header}} \tablewidth{0pt} \tablehead{
 \colhead{Col}
&\colhead{Format}
&\colhead{Units}
&\colhead{Label}
&\colhead{Description}
}\startdata
    1 & A18   & \nodata  & Name  &  SDSS J  \\
    2 & F11.6 & deg   & RAdeg &  Right Ascension in decimal degrees (J2000) \\
    3 & F11.6 & deg   & DEdeg &  Declination in decimal degrees (J2000)\\
    4 & F7.5  & \nodata   & z     &	 redshift\\
    5 & F7.4  & \nodata   & imag  &  PSF i-band apparent magnitude\\
    6 & F7.4  & \nodata   & iMag  &  PSF i-band absolute magnitude,K-corrected to z=2 \\
    7 & I6    & d     & MJD   &  Modified Julian Date \\
    8 & I5    & \nodata   & Plate &  Plate number \\
    9 & I5    & \nodata   & Fiber &  Fiber identification number \\
    10 & F7.4  & {\AA}    & si    &  Instrumental sigma\\
    \hline
    &  &  &  &  Before applying PCA(2) and RPC(2)\\
    \hline
    11 & F7.4  & {\AA}   & sim\_b & $\sigma_{MgII}$ estimates from the second moments \\
    12 & F7.4  & {\AA}    & st\_b & true $\sigma_{MgII}$ in {\AA}  \\
    13 & F9.6  & erg s$^{-1}$ cm$^{-2}$ {\AA}$^{-1}$ & F3\_b & $F_{\lambda}$ at 3000 {\AA} \\
    14 & F7.4  & erg s$^{-1}$ & LL\_b & Monochromatic luminosity at 3000{\AA}, $\lambda L_{\lambda}$ in units 1E44 (4) \\
    15 & F8.4  & light-days & R\_ld\_b & Radius of the BLR  \\
    16 & F8.3  & \nodata   & SN\_G\_b &  Median SNR in G band \\
    17 & F8.3  & \nodata   & SN\_R\_b &  Median SNR in R band \\
    18 & F8.3  & \nodata   & SN\_I\_b &  Median SNR in I band \\
    19 & F7.4  & $M_{\odot}$ & MBH\_bb &  BH mass in solar units \\
    20 & F7.4  & $M_{\odot}$ & Err\_MBH\_bb & Error in BH mass \\
    \hline
    &    &    &    & After applying PCA(2) but before RPC(2)\\
    \hline
    21 & F7.4  & {\AA}    & sim\_a & $\sigma_{MgII}$ estimates from the second moments(3) \\
    22 & F7.4  & {\AA}    & st\_a & true $\sigma_{MgII}$ in {\AA} (3) \\
    23 & F9.6  & erg s$^{-1}$ cm$^{-2}$ {\AA}$^{-1}$ & F3\_a &  $F_{\lambda}$ at 3000 {\AA} (3) \\
    24 & F7.4  & erg s$^{-1}$ & LL\_a &  Monochromatic luminosity at 3000{\AA}, $\lambda L_{\lambda}$ in units of 1E44 (3)(4) \\
    25 & F8.4  & light-days & R\_ld\_a & Radius of the BLR(3) \\
    26 & F8.3  & \nodata   & SN\_G\_a & Median SNR in G band \\
    27 & F8.3  & \nodata   & SN\_R\_a & Median SNR in R band \\
    28 & F8.3  & \nodata   & SN\_I\_a & Median SNR in I band \\
    29 & F7.4  & $M_{\odot}$ & MBH\_ab & BH mass in solar units (3) \\
    30 & F7.4  & $M_{\odot}$ & Err\_MBH\_ab & Error in BH mass (3) \\
    \hline
    &    &    &    & After applying PCA(2) and RPC(2) \\
    \hline
    31 & F7.4  & $M_{\odot}$ & MBH\_aa & BH mass in solar units  \\
    32 & F7.4  & $M_{\odot}$ & Err\_MBH\_aa & Error in BH mass \\
    33 & F7.4  & \nodata   & z\_RH &  redshift re-estimated by this study \\
    34 & I1    & \nodata   & DR3c  &  In DR3 complete subsample? (0=no, 1=yes) \\
    \enddata
    \tablecomments{(1) Entries reading 0.0000 or 0.0000e+00 indicate that the quantity was not measurable.
    (2) PCA stands for Principal Component Analysis and RPC stands for Radiation Pressure Correction.
    (3) Recommended results.
    (4) Bolometric Luminosity can be estimated using $BC(3000{\AA})$ = 5.15 and the monochromatic luminosity at 3000{\AA} (Shen et al. 2008).
    Until the catalogue is available through the ApJS website it can be downloaded as http://ara.phys.yorku.ca/rafieehall2011.txt}
\end{deluxetable}
%\clearpage

\end{document}